\documentclass[aps,onecolumn,nofootinbib,superscriptaddress,letterpaper,amsmath,amssymb]{revtex4}
\pdfoutput=1

\usepackage{graphicx}  
\usepackage{dcolumn}   
\usepackage{bbm}        
\usepackage{amsmath}
\usepackage{amsfonts}
\usepackage{amssymb}   
\usepackage{epstopdf}
\usepackage{slashed}
\usepackage{hyperref}

\usepackage{multirow}
\usepackage{color}
\usepackage{float}
\usepackage{ragged2e}
\usepackage{subcaption}
\DeclareCaptionJustification{justified}{\justifying}
\usepackage[normalem]{ulem}
\usepackage{verbatim}
\usepackage{xspace}

\usepackage[compat=1.1.0]{tikz-feynman}

\def\gsim{\lower0.5ex\hbox{$\:\buildrel >\over\sim\:$}}
\def\lsim{\lower0.5ex\hbox{$\:\buildrel <\over\sim\:$}}

\newcommand{\be}{\begin{equation}}
\newcommand{\ee}{\end{equation}}
\newcommand{\bea}{\begin{eqnarray}}
\newcommand{\eea}{\end{eqnarray}}

\newcommand{\nbox}{{\,\lower0.9pt\vbox{\hrule \hbox{\vrule height 0.2 cm
\hskip 0.2 cm \vrule height 0.2 cm}\hrule}\,}}

\newskip\zatskip \zatskip=0pt plus0pt minus0pt
\def\matth{\mathsurround=0pt}
\def\lsim{\mathrel{\mathpalette\atversim<}}
\def\gsim{\mathrel{\mathpalette\atversim>}}
\def\sigv{\ifmmode \langle\sigma v\rangle\else $\langle\sigma v\rangle$\fi}
\newskip\zatskip \zatskip=0pt plus0pt minus0pt
\def\matth{\mathsurround=0pt}
\def\lsim{\mathrel{\mathpalette\atversim<}}
\def\gsim{\mathrel{\mathpalette\atversim>}}
\def\atversim#1#2{\lower0.7ex\vbox{\baselineskip\zatskip\lineskip\zatskip
  \lineskiplimit
  0pt\ialign{$\matth#1\hfil##\hfil$\crcr#2\crcr\sim\crcr}}}

\begin{document}

\thispagestyle{empty}

\vspace{0.5in}

\title{Fast and Precise Track Fitting with Machine Learning}
\author{Ryan Miller}
\affiliation{Department of  Computer Science, University of California, Irvine, CA}
\author{Alexander Shmakov}
\affiliation{Department of  Computer Science, University of California, Irvine, CA}
\author{Kyuho Oh}
\affiliation{Department of  Computer Science, University of California, Irvine, CA}
\author{Jiwon Lee}
\affiliation{Department of  Computer Science, University of California, Irvine, CA}
\author{Pierre Baldi}
\affiliation{Department of  Computer Science, University of California, Irvine, CA}
\author{Levi Condren}
\affiliation{Department of Physics \& Astronomy, University of California, Irvine, CA}
\author{Makayla Vessella}
\affiliation{Department of Physics \& Astronomy, University of California, Irvine, CA}
\author{Daniel Whiteson}
\affiliation{Department of Physics \& Astronomy, University of California, Irvine, CA}

\begin{abstract}
Efficient and accurate particle tracking is crucial for measuring Standard Model parameters and searching for new physics. This task consists of two major computational steps: track finding, the identification of a subset of all hits that are due to a single particle; and track fitting, the extraction of crucial parameters such as direction and momenta. Novel solutions to track finding via machine learning have recently been developed. However, track fitting, which traditionally requires searching for the best global solutions across a parameter volume plagued with local minima, has received comparatively little attention.
Here, we propose a novel machine learning solution to track fitting. The per-track optimization task of traditional fitting is transformed into a single learning task optimized in advance to provide constant-time track fitting via direct parameter regression. This approach allows us to optimize directly for the true targets, i.e., the precise and unbiased estimates of the track parameters. This is in contrast to traditional fitting, which optimizes a proxy based on the distance between the track and the hits. In addition, our approach removes the requirement of making simplifying assumptions about the nature of the noise model. Most crucially, in the simulated setting described here, it provides more precise parameter estimates at a computational cost that is more than 1,000 times smaller, which leads directly to improvements in particle momentum estimation, vertex finding, jet-substructure identification, and anomaly detection. 
\end{abstract}
\maketitle

\section{Introduction}
\label{Sec:Introduction}

Particle tracking plays a vital role in the scientific program of particle physics experiments~\cite{ATLAS:2012yve,CMS:2012qbp,annurev:/content/journals/10.1146/annurev-nucl-101920-014923,annurev:/content/journals/10.1146/annurev-nucl-102419-052854} in collider, fixed-target and astro-particle contexts, by providing precise information about particle trajectories. Tracking information is essential for estimating particle momentum~\cite{ATLAS:2020auj,CMS:2018rym}, establishing location of vertices~\cite{ATL-PHYS-PUB-2019-015,ATL-PHYS-PUB-2019-013}, extracting substructure from showers~\cite{ATLAS:2018nuj,Lee:2023xzv}, and identifying the nature of deposits in subsequent detectors~\cite{ATL-PHYS-PUB-2017-011, electronreco}, among other applications.  

The tracking task has two distinct steps, finding and fitting. Track finding identifies a set of detector hits that are due to a single particle, among the many hits in an event. Track fitting is the extraction of particle parameters by fitting to a particular hypothesis, such as a helical trajectory for particles in a magnetic field.   The first step, track finding, presents an enormous combinatorial challenge, especially with the increased hit density expected with a higher instantaneous luminosity~\cite{run3cpuperf}. Novel machine learning based approaches have recently been brought to bear~\cite{Bronstein:2016thv,ExaTrkX:2020nyf,ExaTrkX:2021abe} with greater performance and flexibility, including the natural generalization to non-helical tracks~\cite{Kang:2008ea, Sha:2024hzq}. However, the fitting problem has received significantly less attention.

Traditional approaches to track fitting use maximum likelihood estimation to fit a parametric function. This requires making assumptions about the hit uncertainty and searching a vast space of parameters with many local minima 
using heuristic optimization algorithms.  Such fitters are famously sensitive to the choice of initial search parameters and can be unpredictable and computationally expensive.  Compounding these issues,  some track finding algorithms run track fitting many times to refine the search region~\cite{10.1115/1.3662552}. High-precision track fitting is  computationally expensive, yet tracking is important in situations where speed is essential. Most previous approaches to fast track fitting have relied on lookup table approaches~\cite{Ohsugi:1988rh}, which sacrifice precision and memory for speed, or on linearization approaches~\cite{ATLAS:2021tfo}, which solve an approximate version of the problem, again sacrificing precision. A recent machine-learning approach~\cite{Alonso-Monsalve:2022zlm} in the context of high-fidelity neutrino experiments explored using ML to improve iterative hit selection over the standard propagation techniques.

In this paper, we develop and evaluate a fast and precise machine-learning-based track fitter. The burden and unpredictability of the per-track optimization task in traditional fitting is transformed into a single learning task,  optimized in advance to provide reliably and precise fast fitting via direct parameter regression.  This approach allows us to optimize directly for the true targets, precise and unbiased estimates of the track parameters, rather than for the proxy used in traditional fitting, based on the distances from the track to the hits. It removes the requirement, present in traditional fitting algorithms, of making simplifying assumptions about the nature of the noise. Most crucially, it provides  precise track parameter estimates  at a computational cost 1,000 times smaller.  Given the wide application of tracking, including for example particle identification, particle reconstruction, vertex finding, jet substructure measurements and anomaly detection, a boost in precision track fitting performance can have broad impacts on the physics program of particle physics experiments. 

This paper is organized as follows. Section~\ref{sec:bg} describes the track fitting task in formal terms and contrasts our approach with the traditional strategy. Section~\ref{sec:samp} describes the simulated tracks used for testing and training. Section~\ref{sec:tracking} describes our implementation of the traditional least-squares method as well as the deep learning \cite{baldi2021deep} model structure and training. Section~\ref{sec:perf} compares the performance of the two methods. Section~\ref{sec:ad} explores the impact in the context of track anomaly detection, and Section~\ref{sec:conc} summarizes the results and describes future directions.

\section{Background}
\label{sec:bg}

Electrically charged particles move in a helical trajectory in a solenoidal magnetic field $B$, with radius $r= mv/qB$, for particle with mass $m$, velocity $v$, and charge $q$. The helical path can be parameterized in five dimensions:
\[ \bar{p} = (\phi_0, p_\textrm{T}, d_0, d_z, \tan(\lambda)) \]

\noindent which are the initial angle ($\phi_0$), the transverse momentum ($p_\textrm{T}$), the distance of closest approach to the origin in the transverse plane ($d_0$), the closest approach to the origin along the $z$ axis ($d_z$) and the slope in the $rz$ plane ($\tan(\lambda)$).  These are the parameters of interest which allow, for example, the calculation of particle momentum or the identification of vertices. 

\begin{figure}
    \centering
    \includegraphics[width=0.55\linewidth]{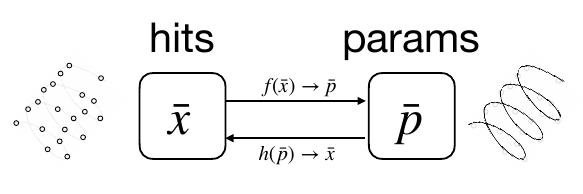}
    \caption{The hit space describes the location of intersections of the track with the detector layers: $\bar{x} = \{x_i\}$. The parameter space describes the five parameters of the helix $\bar{p} = (\phi_0,p_\textrm{T},d_0,d_z,\tan(\lambda))$. The analytic helix function
$h(\bar{p})$ maps the parameters to the hits, while the unknown  function $f(\bar{x})$ maps the hits to the parameters.}
    \label{fig:spaces}
\end{figure}

Where the path of the particle intersects tracking detector layers, a hit can be registered with some finite resolution. The detector layers are located at fixed radii away from the center, producing three-dimensional spatial observations at each radius. Track {\it finding} is the task of identifying the set of hits left by a particle. The challenge of track {\it fitting} is to recover physical parameters $\bar{p}_0$ of the particle responsible for the detector hits $\bar{x}_0$.  While there is a simple analytical expression to map helical parameters to hits $h(\bar{p}) \rightarrow \bar{x}$, the inverse function $f(\bar{x}) \rightarrow \bar{p}$ has no simple closed-form expression; see Fig.~\ref{fig:spaces}.

In the absence of $f(\bar{x})$, traditional track fitting  takes aim at a closely related task: given a set of hits $\bar{x}_0$ with uncertainty $\sigma$, find $\bar{p}$ to minimize:

\[ \chi^2_{\textrm{hit}} =\frac{[h(\bar{p}) - \bar{x}_0]^2}{\sigma^2} = \frac{ [\bar{x} - \bar{x}_0]^2}{\sigma^2} \]

Least-squares fitting  directly minimizes this hit-to-fitted-track distance, rather than the residual in the track parameters $\bar{p}-\bar{p}_0$.  If one assumes the noise model $\sigma$ is Gaussian, this minimization strategy is equivalent to maximum likelihood estimation where the likelihood is $L(\bar{p} | \bar{x}_0) = G[\bar{x}_0,\mu = h(\bar{p}), \sigma]$.  Under these assumptions, minimization of $\chi^2_{\textrm{hit}}$ will also tend to minimize $\bar{p}-\bar{p}_0$, providing an unbiased estimator of the track parameters. Note that expected variance of the individual parameters is linked by their particular relationships to $\chi^2_{\textrm{hit}}$ via $h(\bar{p})$, which can be highly non-linear and can favor larger residuals in one parameter over another to minimize $\chi^2_{\textrm{hit}}$, such that minimizing the $\chi^2_{\textrm{hit}}$ does not guarantee minimal variance in any particular track parameter.

\begin{figure}
    \centering
    \includegraphics[width=0.5\linewidth]{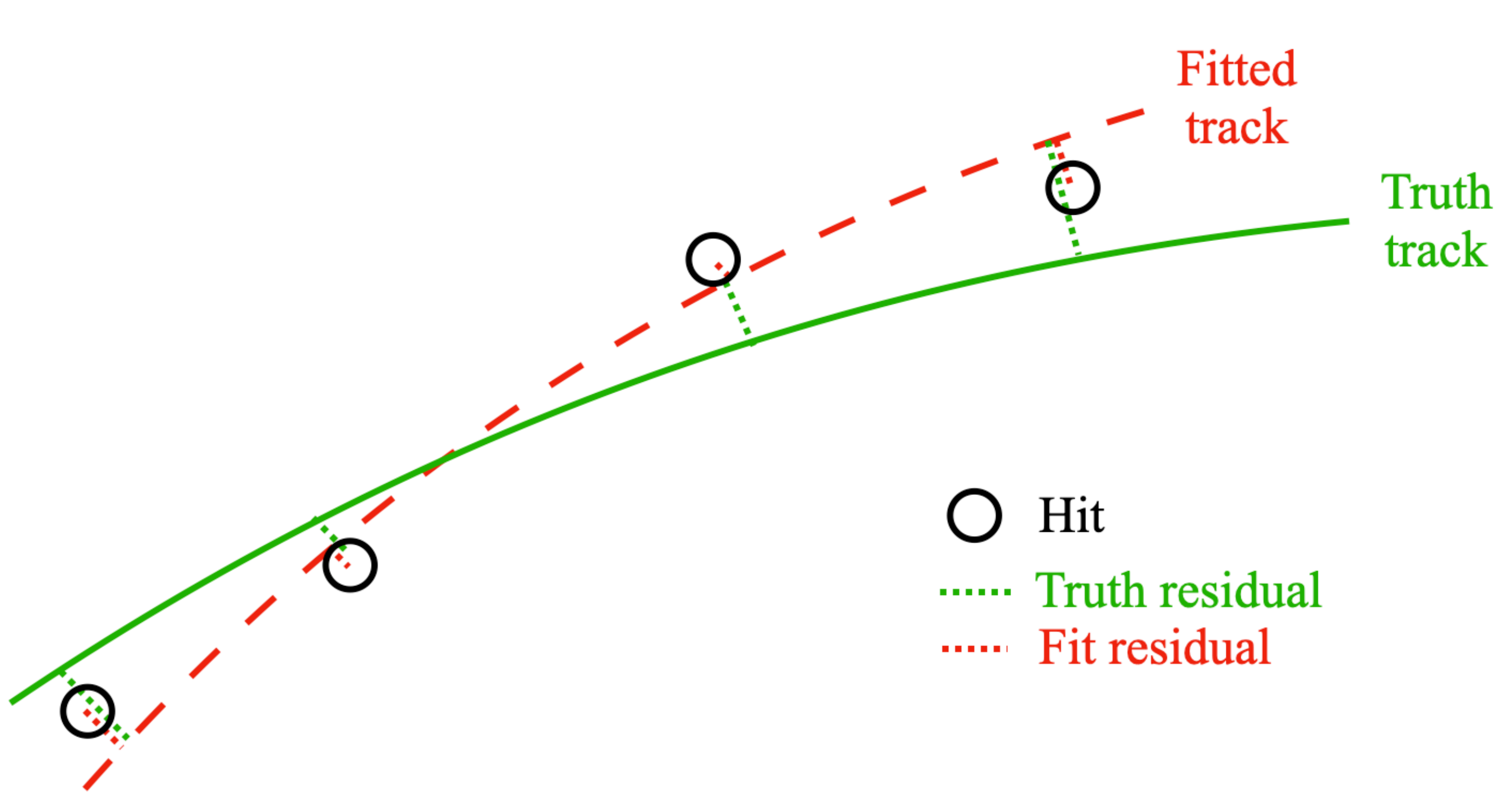}
    \caption{ Traditional tracking seeks to minimize the distance between the hits (black circles) and the fitted track (dashed red), the hit residual. This is often an excellent proxy for estimation of the parameters of the truth track (green) by minimizing the truth residual.}
    \label{fig:defs}
\end{figure}

 Advantages of the least-squares approach are that $\chi^2_{\textrm{hit}}$ can be calculated per-track with  no truth information ({\it i.e.} $\bar{p}_0$), and so can used for fitting and goodness-of-fit testing of tracks in data. In the following, we set $\sigma=1$ for simplicity of notation and refer to this quantity, $\chi^2_{\textrm{hit}}$, as the {\it hit residual}, a measure of the distance between the fitted track and the hits, to distinguish it from the true target {\it truth residual}, a measure of the distance between the fitted track and  true track parameters. See Fig.~\ref{fig:defs} for a visual depiction.

The challenges of the least-squares approach are that it requires scanning the space of $\bar{p}$, which has many local minima. Searching the space is therefore non-trivial, requiring heuristic optimization which is sensitive to the quality of the initial guesses.   In addition, minimization of $\chi^2_{\textrm{hit}}$ is only a valid proxy for the true task, minimization of $\bar{p}-\bar{p}_0$, when the assumption of Gaussian noise holds. While many real-world sources of hit noise are approximately Gaussian, this assumption breaks down with varying impact in cases such as multiple-scattering effects at low momentum or Bremsstrahlung radiation.  Even when it holds, minimizing the scalar $\chi^2_{\textrm{hit}}$ introduces an implicit trade-off between the value of precision in each track parameter, as equal shifts in the parameters do not yield equal penalty in $\chi^2_{\textrm{hit}}$; see App~\ref{app:chisq}. This forces the least-squares approach to make larger sacrifices in precision for some parameters in exchange for smaller improvements in others. 

Here, instead, we seek to recover the unknown $f(\bar{x}) \rightarrow \bar{p}$ that minimizes $\bar{p}-\bar{p}_0$, posing it as a machine learning regression task to overcome the lack of a simple closed-form expression.   This allows for optimization for the true target of track parameter precision, rather than an approximate proxy. It removes the requirement of explicitly specifying a noise model, and any assumptions or constraints it must satisfy to ensure fit convergence, in favor of an implicit noise model defined by the training sample, allowing for a broader set of possibilities, including noise models extracted from as-built detectors in test beam data. It provides a constant-time result, avoiding the computationally expensive and unreliable search through the poorly-behaved space of $\bar{p}$. It requires knowledge of the value of the track parameters $p_0$, but only during the training phase to find $f(\bar{x})$.

\section{Data}
\label{sec:samp}

\begin{figure}
    \includegraphics[width=0.6\linewidth]{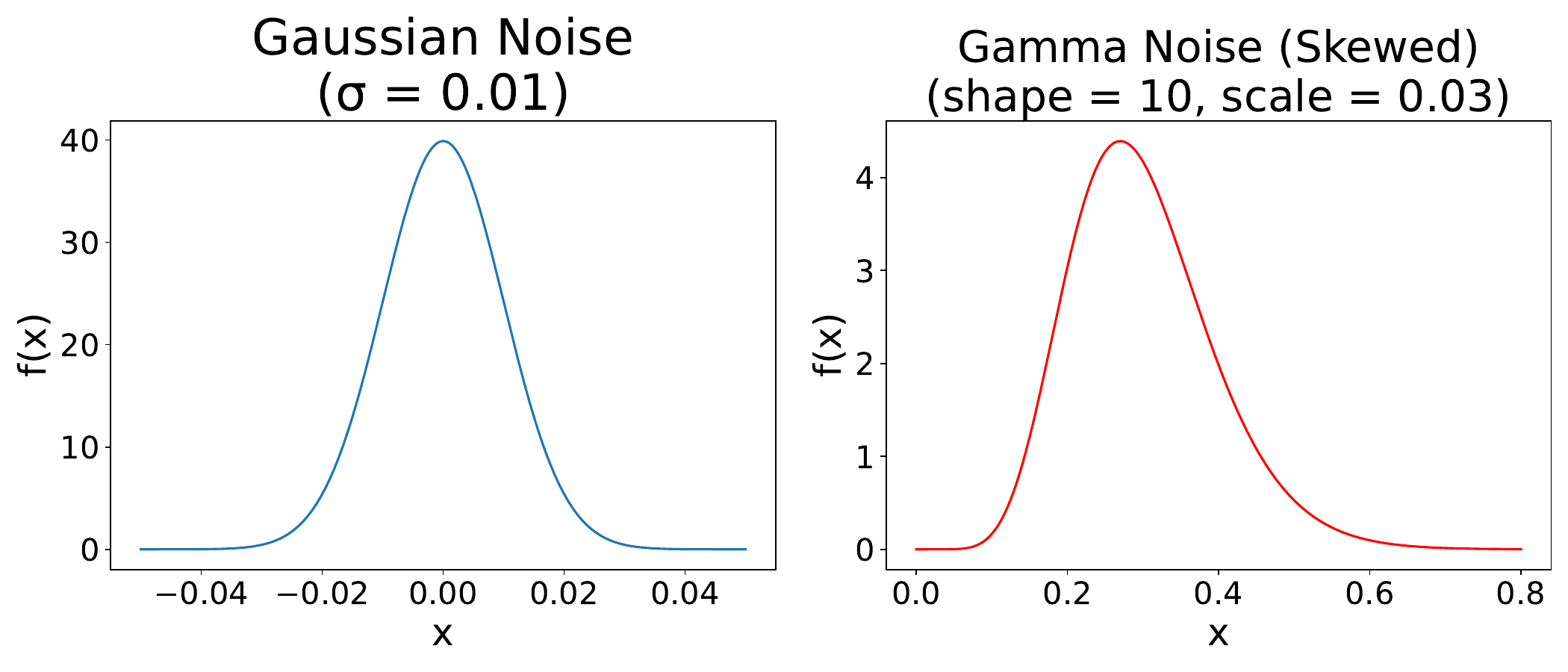}
    \caption{Track positions are smeared using one of two noise models, either a Gaussian (left) centered at the true location of the detector intercept, or a Gamma function (right) which introduces a bias.}
    \label{fig:noise_models}
\end{figure}

Samples of simulated tracks are generated for training our regression model, as well as for evaluating the performance of the model and the benchmark least-squares fitter. 
Tracks are generated with a range of parameter values; see Tab.~\ref{tab:track_params} for details. Once the parameters $\bar{p}_0$ are specified, the track's path through space is described using a standard helix parameterization~\cite{helix}:

\[ x(t) = d_0 \cos(\phi_0) + \frac{p_\textrm{T}}{qcB}[\cos(\phi_0) - \cos(\phi_0 + t)]  \]
\[ y(t) = d_0 \sin(\phi_0) + \frac{p_\textrm{T}}{qcB}[\sin(\phi_0) - \sin(\phi_0 + t)]  \]
\[ z(t) = d_z - \frac{p_\textrm{T}}{qcB} \tan(\lambda) t\]

\noindent where $q$ is the charge and $B$ is the magnetic field strength. The set of hits are defined as the intersection of the path with the detector layers. For this proof-of-concept study, we use a simple geometry, consisting of ten cylindrical layers at radii $\{1,2,...10\}$ cm. Observations are generated by adding independent random noise $\epsilon$ to each coordinate. This noise is sampled from either a centered Gaussian, $\mathcal{N}(\mu = 0, \sigma = 0.01)$ or a Gamma distribution, $\Gamma(\alpha=10, \theta=0.03)$, whose peak is offset from the true location (Fig.~\ref{fig:noise_models}).

\begin{figure}
    \includegraphics[width=0.45\linewidth]{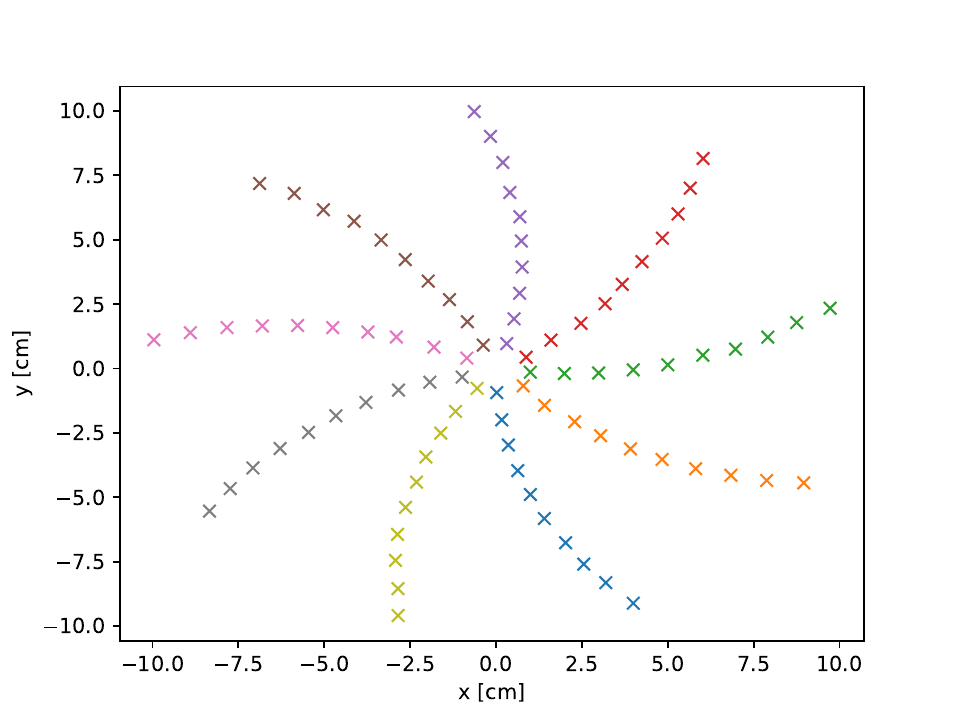}
    \includegraphics[width=0.45\linewidth]{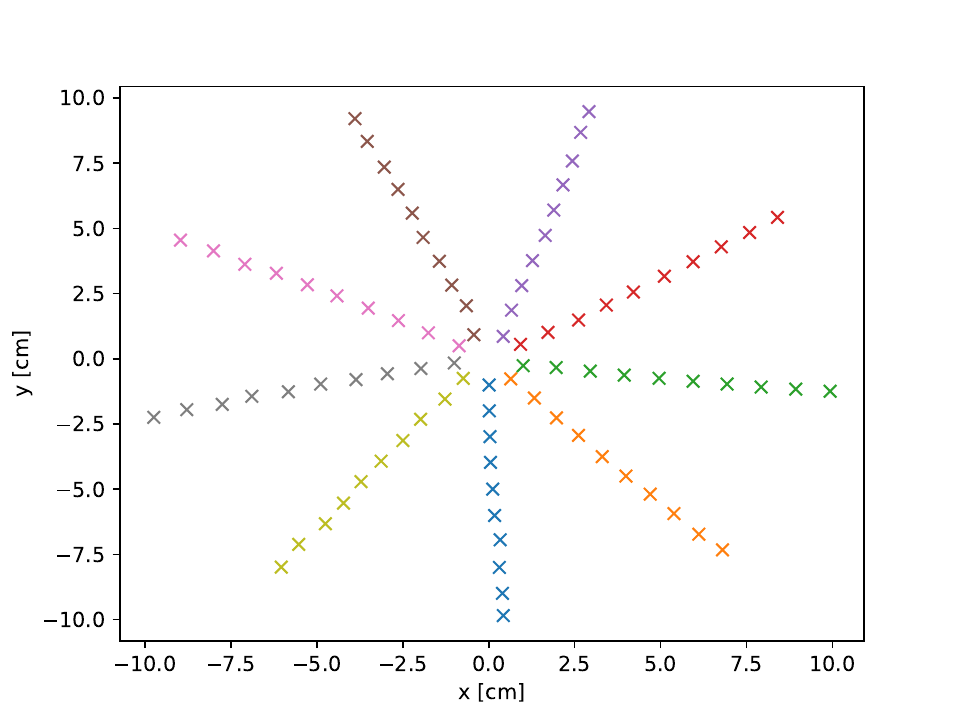}
        \caption{Examples of helical generated tracks with varying initial angle for two values of $p_\textrm{T}$, 25 GeV (left) or 200 GeV (right).}
        \label{fig:gentracks}
\end{figure}

We generate 1M tracks and perform an 80/10/10 train/val/test split. Selected examples are shown in Fig.~\ref{fig:gentracks}.

\begin{table}[htbp]
  \centering
  \begin{tabular}{lcc}
    \hline
    \textbf{Parameter} & \textbf{Distribution} & \textbf{Range} \\
    \hline
    $d_0$   & Half-Normal ($\sigma=0.03$)  & $[0,\infty)$ cm \\
    $\phi$  & Uniform   & $[0,2\pi]$ rad  \\
    $p_\textrm{T}$   & Log-Normal ($\mu=4, \sigma=0.5$)   & $(0,\infty)$ GeV  \\
    $dz$    & Normal ($\sigma=0.5$) & $(-\infty,\infty)$ cm\\
    $\tan(\lambda)$ & Normal ($\sigma=0.6$) & $(-\infty,\infty)$ \\
    \hline
  \end{tabular}
  \caption{Ranges and distributions of the helical track parameters.}
  \label{tab:track_params}
\end{table}

\section{Track Fitting}
\label{sec:tracking}
Track fitting is done via traditional least-squares fitting, to serve as a performance benchmark, as well as via neural network regression.

\subsection*{Least Squares Fitting}

To determine the best-fit helical parameters for each particle trajectory, we employ a multi-stage fitting procedure that balances computational efficiency with accuracy. First, we generate an initial estimate of the track parameters using a fast, approximate method~\cite{HANSROUL1988498}.

Parameter refinement is performed by minimizing $\chi^2_\textrm{hit}$, as defined above, which measures the distances between the predicted and measured hit positions. We use the L-BFGS-B algorithm\cite{liu1989limited}, as implemented in \texttt{scipy.optimize.minimize}\cite{2020SciPy-NMeth}, with parameter bounds 5 standard deviations away from true parameter distribution means. These bounds are adaptive: if the initial helical parameter guess lies outside the global range, the corresponding bound is extended to the initial guess.

Before applying this minimization to the final fit, we explore the local neighborhood of the fast-fit parameters using coarse two-dimensional grid searches in orthogonal subspaces — first in the $(\phi_0,1/p_T)$ plane (XY projection) and then in the $(d_z,\tan\lambda)$ plane (Z projection). For each trial point in these grids, we run a bounded L-BFGS-B minimization to evaluate and refine the parameters, keeping the best parameters found so far. This targeted search steers the optimizer toward more favorable basins and improves the starting point for the final fit.

Once the grid search phase is complete, we perform a final L-BFGS-B refinement starting from the best parameters found. If the fit quality remains poor (large $\chi^2_\textrm{hit}$), we repeat the refinement several times from perturbed versions of the current best parameters, allowing the procedure to escape shallow local minima.

This combination of fast initialization, and constrained iterative refinement reduces sensitivity to poor starting points and avoids the computational cost of a full-scale unconstrained search. This method serves as our benchmark, against which we compare the performance of our ML approach.

This version of a LS fitter is not guaranteed to find the global minimum and produce the ideal fit, reflecting the challenge facing users of traditional track fitters who must balance the expense of more aggressive optimization with the need for accurate fitting.   To separate these issues, and provide as robust a benchmark comparison as possible, we provide a second, idealized version of the LS fitter, in which the optimization begins in the immediate neighborhood of the true values. Such a fitter requires information not usually present during track fitting, so cannot be used in experiments, but gives an upper bound on what might be achieved through extremely effective optimization.


\begin{figure}
    \centering
    \includegraphics[width=0.9\linewidth]{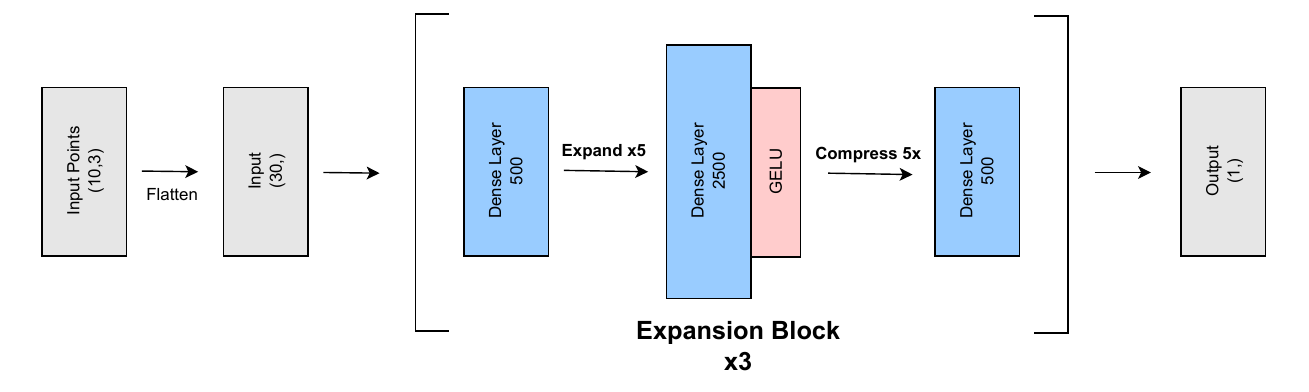}
    \caption{ Structure of the model, which learns to map the input data (10 3-dimensional space points) to one of the helix parameters.}
    \label{fig:model}
\end{figure}

\subsection*{Neural Network Regression}

To predict the helical track parameters, we adopt a committee-of-experts approach~\cite{COE}, training a separate neural network (NN) for each individual parameter rather than using a single multi-output model. Each network follows a multi-layer perceptron (MLP) architecture with 3 expansion-contraction blocks of size 500 and an expansion factor of 5, inspired by inverted-bottleneck layers introduced in \cite{liu2022convnet, Sandler_2018_CVPR}. The specific layer dimensions are shown in Fig.~\ref{fig:model}. The input to each network is a flattened vector of 10 three-dimensional points, yielding 30 input features, and the output is a single scalar corresponding to one of the helical parameters. We initialize all weights using He normal initialization\cite{datta2020surveyactivationfunctionsrelation}, which is well-suited for GELU activations and helps stabilize deep network training. Input coordinates were left unnormalized, as normalization did not lead to any performance improvement.

To handle the angular parameter $\phi_0$, we train two separate models to predict $\sin(\phi_0)$ and $\cos(\phi_0)$ instead of $\phi_0$ directly. When evaluating on the test dataset, we recompute the value of $\phi_0$ by computing the arc-tangent of $\sin(\phi_0)/\cos(\phi_0)$. We then reduce the reconstructed $\phi_0$ modulo $2\pi$ to map the predicted $\phi_0$ into the domain of $[0,2\pi)$. This avoids discontinuities at the $2\pi$ boundary and results in more stable learning behavior. Empirically, we observed significantly improved performance using this formulation. In total, we train six models—one for each scalar output, including the two for $\phi_0$.

Each model is trained independently using the Adam optimizer~\cite{kingma2017adammethodstochasticoptimization} with a learning rate of 0.0001, $\beta_1=0.9$, $\beta_2=0.99$, and mean squared error (MSE) as the loss function. Regularization techniques such as dropout~\cite{10.5555/2627435.2670313} and weight decay~\cite{loshchilov2019decoupledweightdecayregularization} reduced performance, and were not used. Training is conducted over 1000 epochs with a batch size of 4096. This committee-of-experts strategy enables each model to specialize in predicting a single parameter, and leads to improved performance at the cost of increased computational overhead.

In our approach, we view the mixture of experts model as a form of regularization, where we impose a block-diagonal structure on every weight matrix, effectively decomposing the network into independent experts that each specialize in a single helix parameter. We do this based on the observation that the five helix parameters are modestly correlated: knowing one parameter provides limited predictive power over the others. While correlations between parameters do exist in real-world detector scenarios, for example between $d_{0}$ and $dz$ in displaced tracks, in practice this decomposition provides a reasonable first-order description.
By assigning each helix parameter to its own expert, we reduce overall complexity and recast the inversion into six focused subproblems—trading off capacity to model inter-parameter interactions in exchange for markedly more stable convergence and higher accuracy. A unified network, by contrast, must capture all of the helix’s highly non-linear relationships at once, which demands far greater capacity and may lead to unstable training and overfitting. Although such a unified model should exist in principle, additional work is required to close the performance gap with our committee of experts approach. In practice, the mixture-of-experts approach offers an efficient and effective approximation to the full inversion task. 

\section{Performance}
\label{sec:perf}

Performance of each approach, under both noise models, is assessed primarily by evaluating the track parameter residuals, $\bar{p}-\bar{p}_0$.  Residuals with mean close to zero indicate a lack of bias for extracting physical parameters, while residuals with small variance indicate small uncertainties. Computational cost is also measured, with all model evaluation performed on an Intel(R) Xeon(R) CPU E5-2620 v3 processor.

\begin{figure}[h!]
    \centering
    \includegraphics[width=0.3\linewidth]{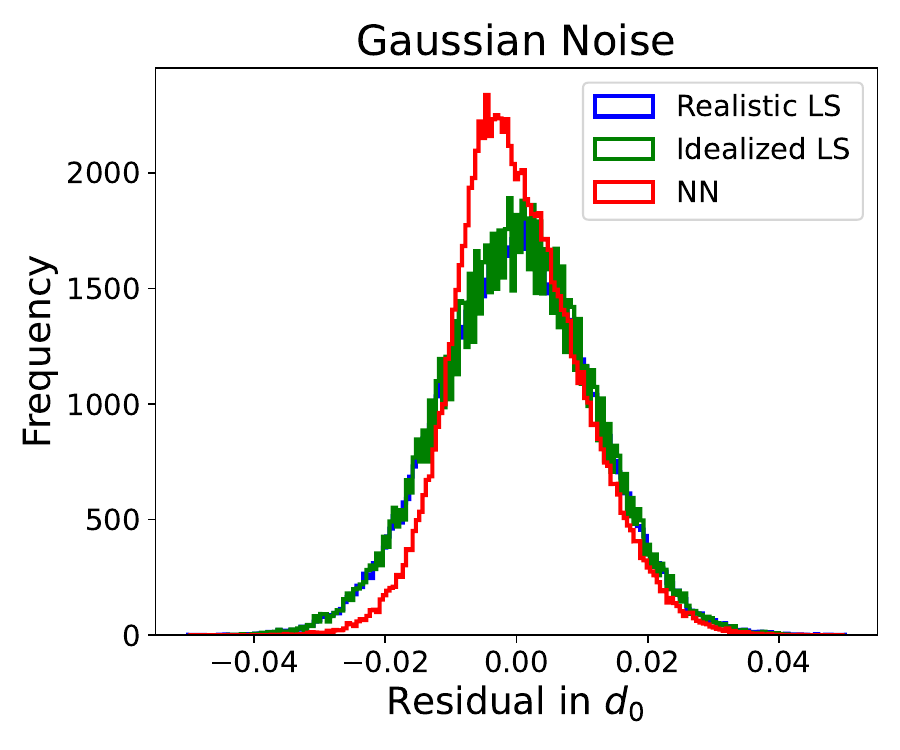}
     \includegraphics[width=0.3\linewidth]{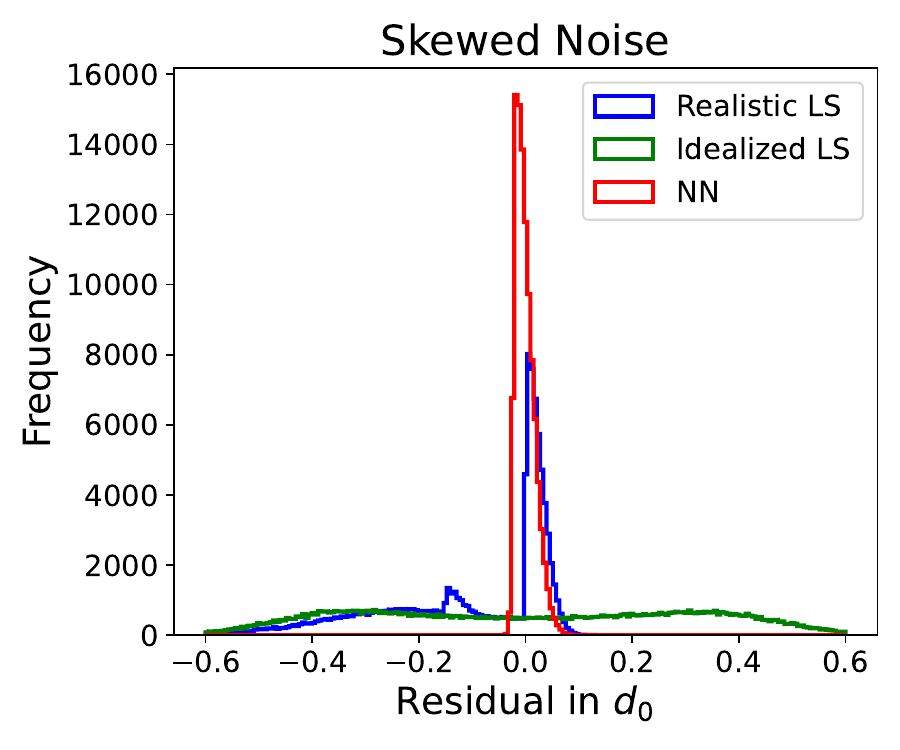}\\
    \includegraphics[width=0.3\linewidth]{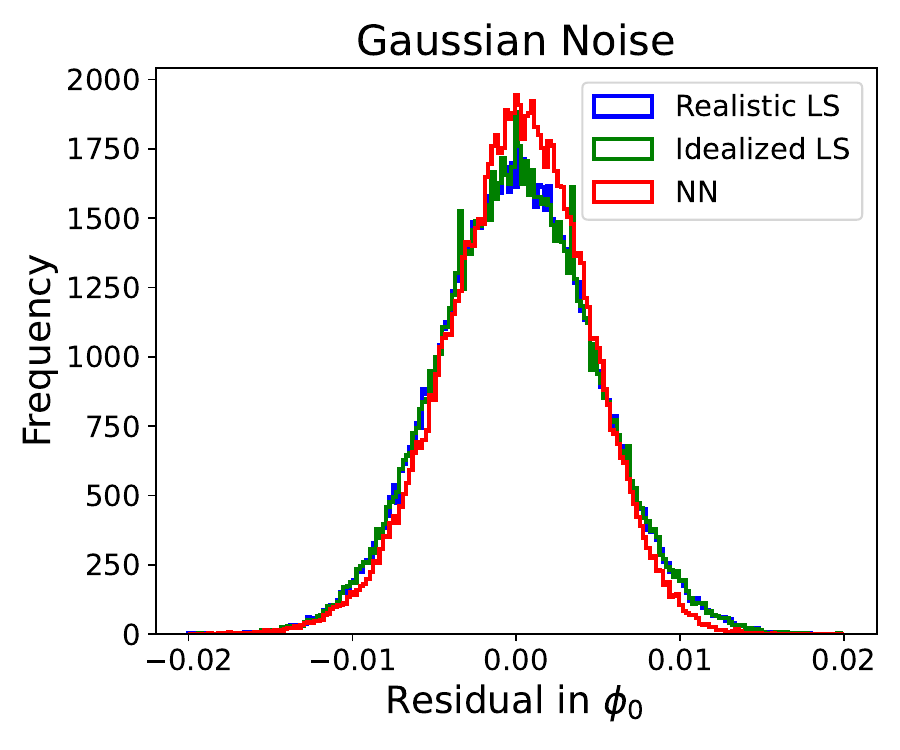}
        \includegraphics[width=0.3\linewidth]{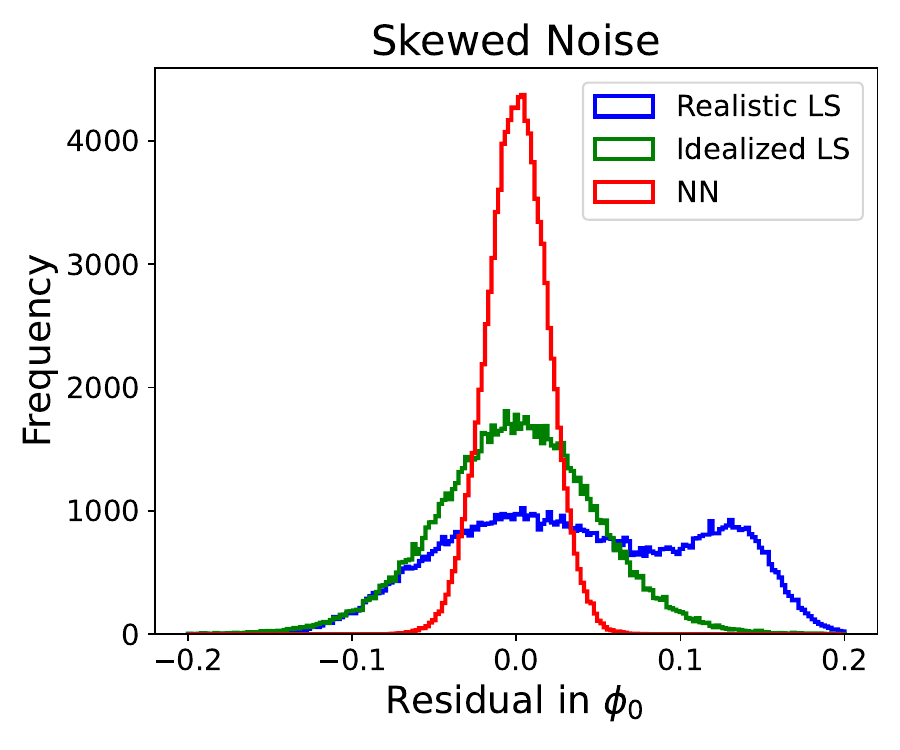}\\
    \includegraphics[width=0.3\linewidth]{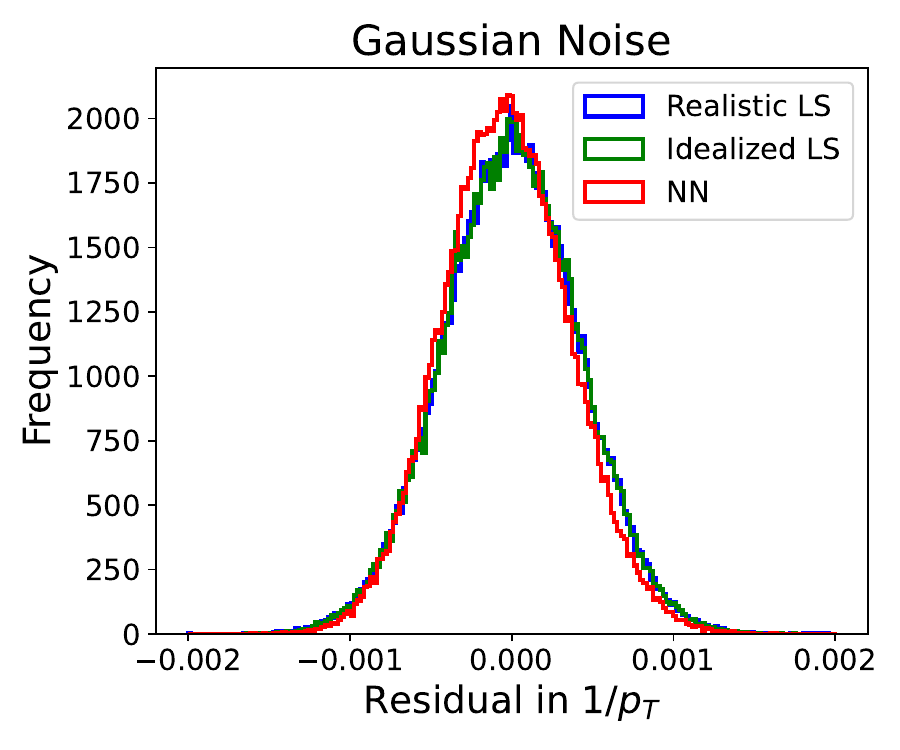}
        \includegraphics[width=0.3\linewidth]{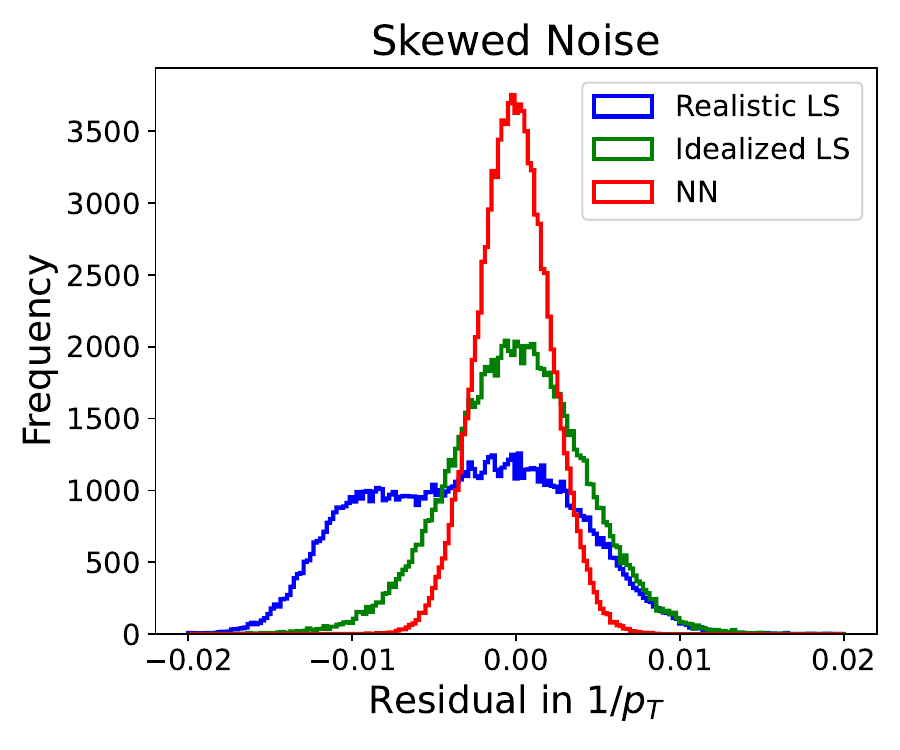}
    \caption{Performance of track parameter regression for NN model, traditional LS fitting and an idealized LS fitter, in terms of track parameter residuals $d_0,\phi_0, 1/p_{\textrm{T}}$.  Left: Gaussian hit noise, right: skewed hit noise model.}
    \label{fig:residuals}
\end{figure}

\begin{figure}[h!]
    \centering
    \includegraphics[width=0.3\linewidth]{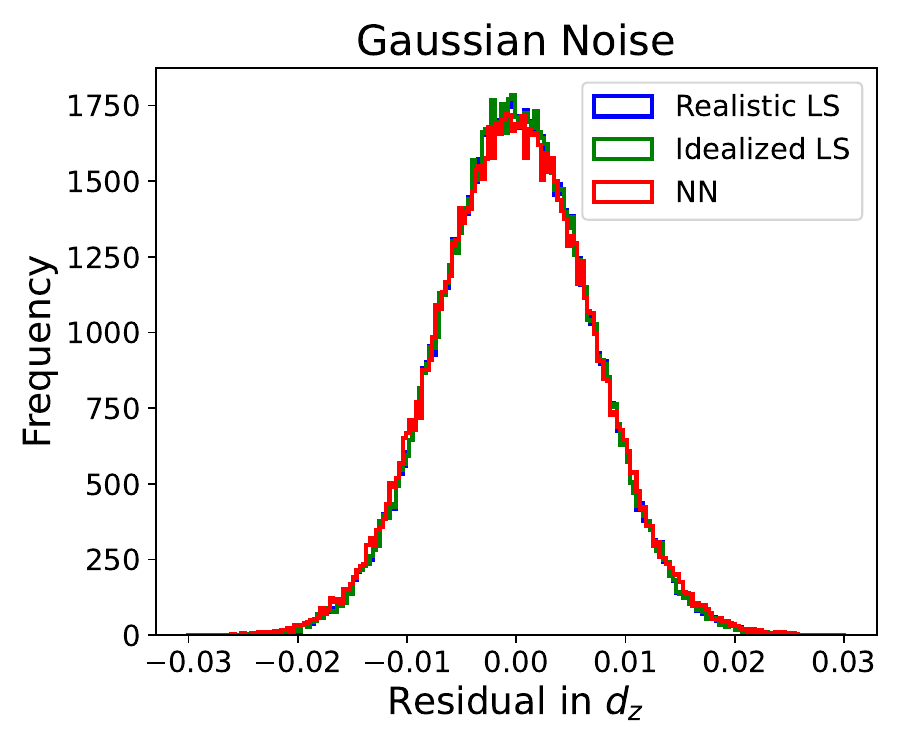}
     \includegraphics[width=0.3\linewidth]{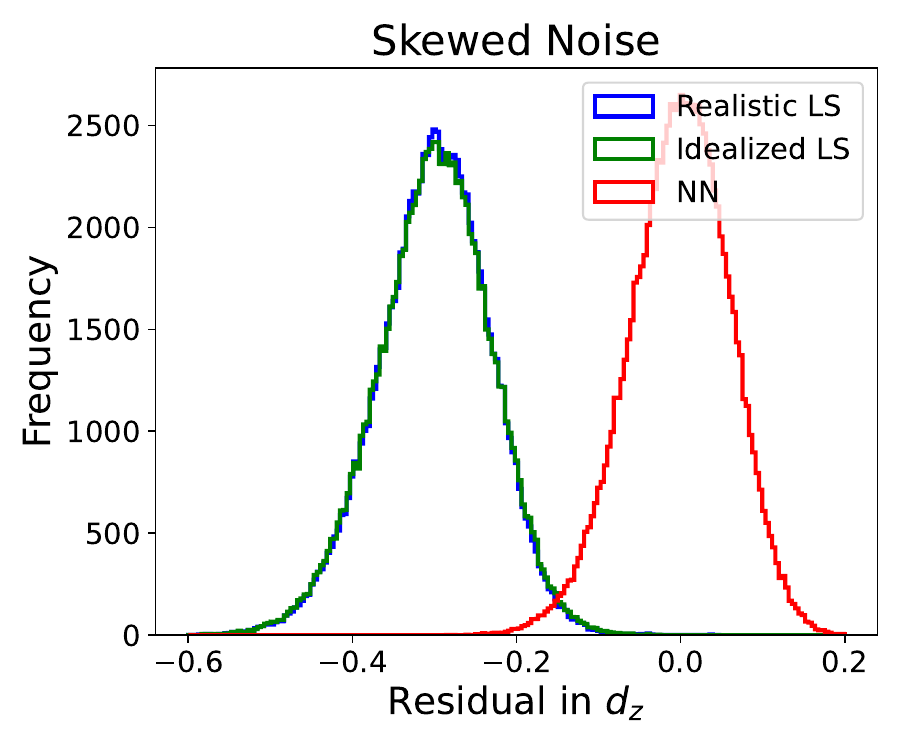}\\
    \includegraphics[width=0.3\linewidth]{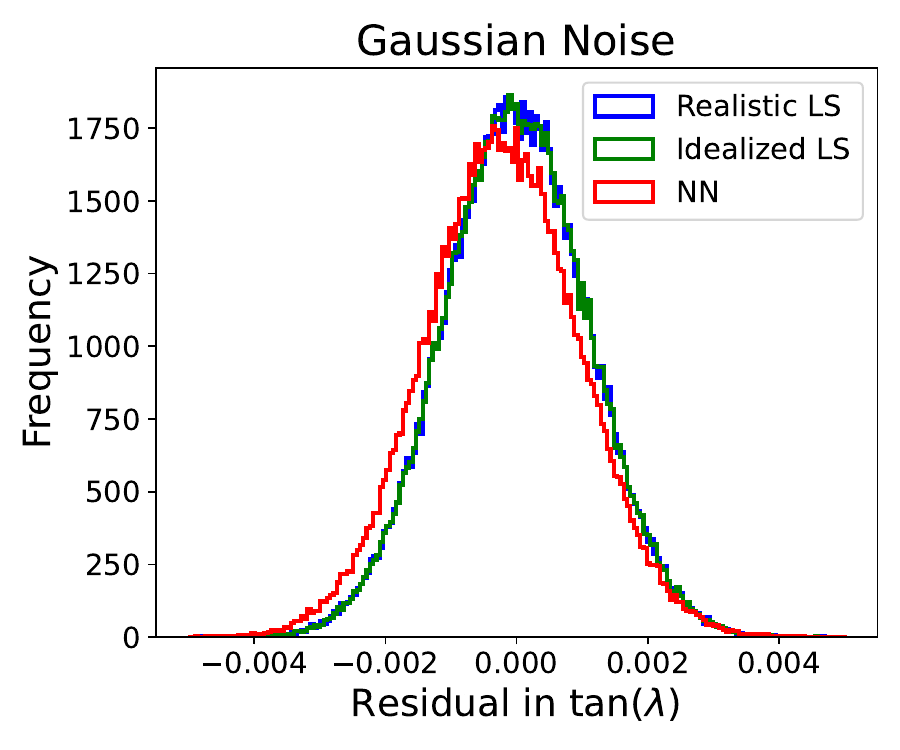}
        \includegraphics[width=0.3\linewidth]{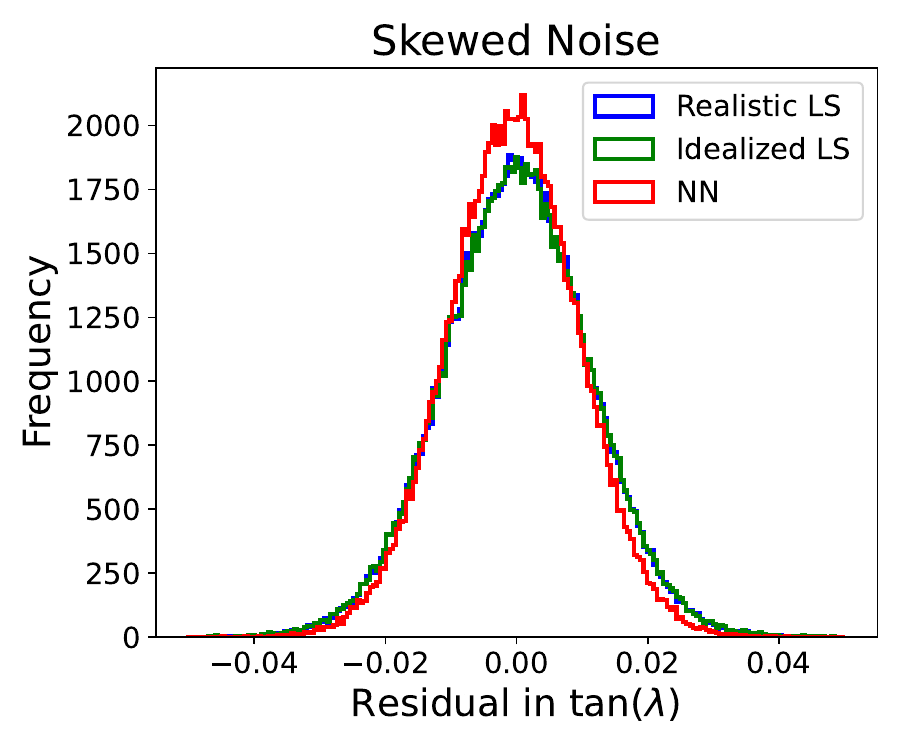}
    \caption{Performance of track parameter regression for NN model, traditional LS fitting and an idealized LS fitter, in terms of track parameter residuals $d_z,\tan(\lambda)$.  Left: Gaussian hit noise, right: skewed hit noise model.}
    \label{fig:residualsz}
\end{figure}

\begin{figure}[h!]
    \centering
    \includegraphics[width=0.4\linewidth]{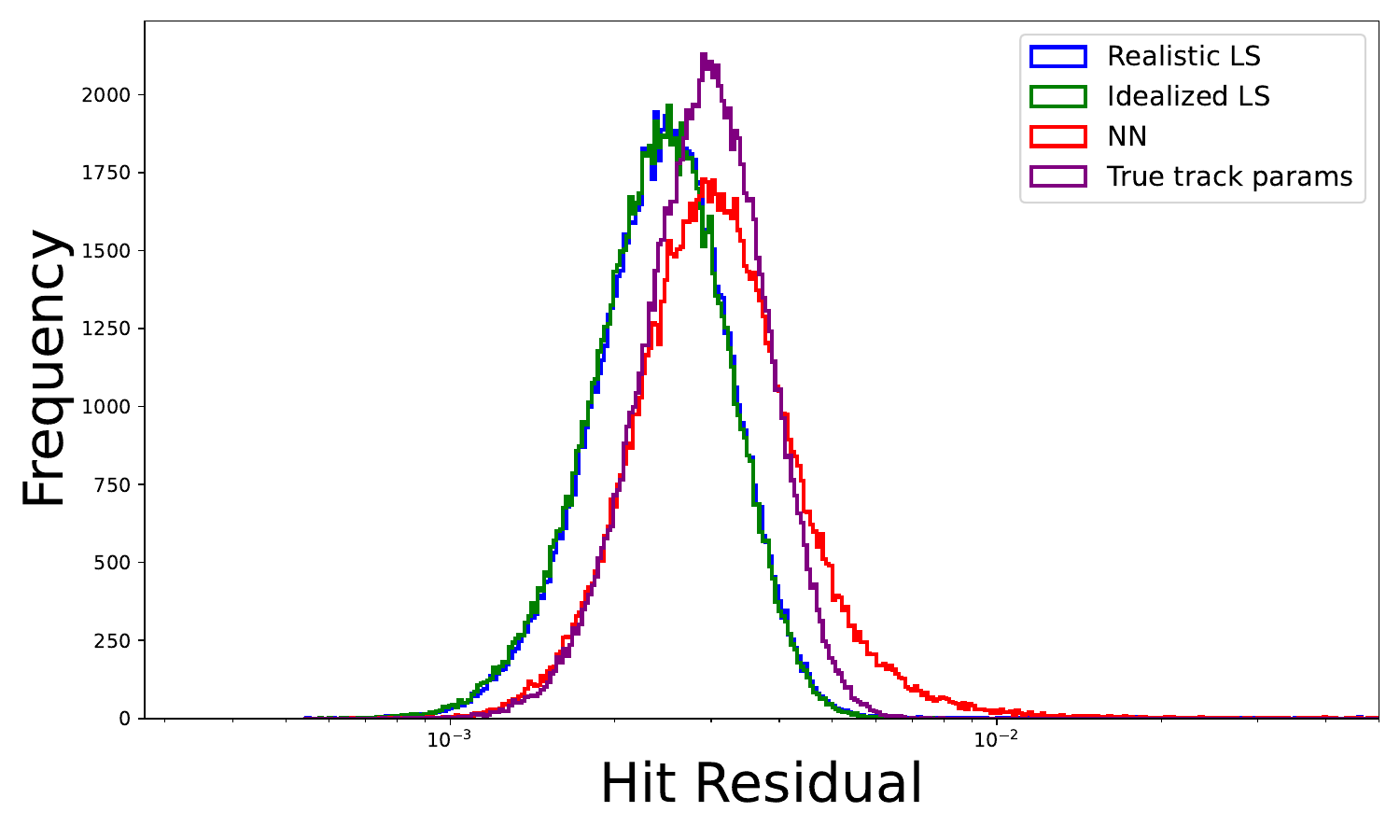} 
    \includegraphics[width=0.4\linewidth]{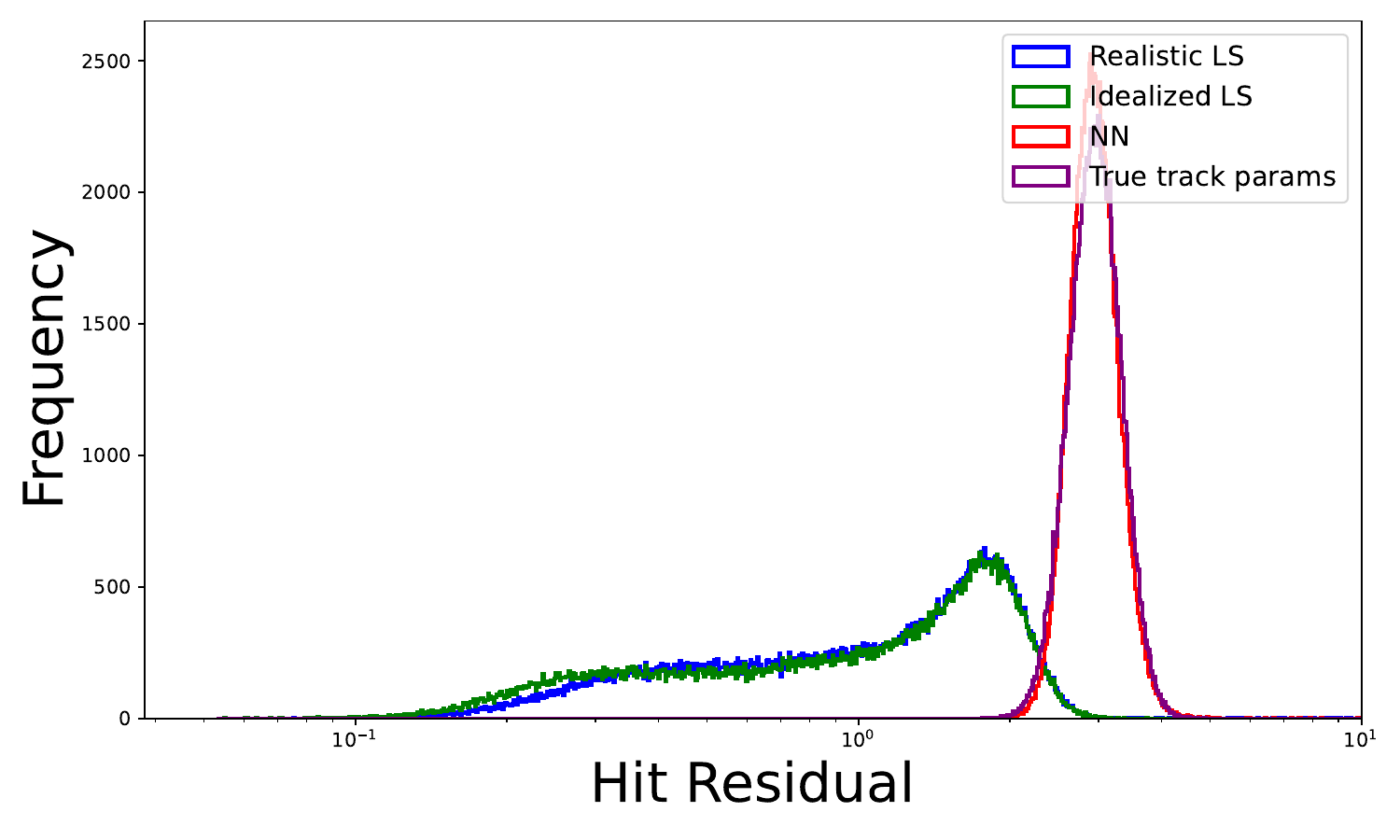}
    \caption{Distribution of the $\chi^2_\textrm{hit}$ metric for NN and LS fitters. Left, for Gaussian noise; right, for skewed.}
    \label{fig:chisq}
\end{figure}

Distributions of the individual parameter residuals for algorithms and each noise model are shown in Figs.~\ref{fig:residuals} and ~\ref{fig:residualsz}.  Means and variances are presented in Tab.~\ref{tab:helical_parameters}.  The traditional hit residual, $\chi^2_\textrm{hit}$, which measures the fitted-track-to-hit distance
is shown in Fig.~\ref{fig:chisq}.  The NN is able to predict track parameters at a speed of $2.08 \times 10^{-4}$ s/track using a batch size of 100,000, while the LS method predicts tracks at a speed of 0.268 s/track, a relative factor of over 1,000. If the NN models are run on GPU, this factor can be increased significantly further in favor of the NN.

\begin{table}[h!]
    \centering
    \caption{Mean absolute errors with standard deviations of helical parameters for Gaussian and skewed datasets.}
    \label{tab:helical_parameters}
    \begin{tabular}{c|ccc|ccc}
        \hline\hline
        \multirow{2}{*}{\textbf{Parameter}} 
        & \multicolumn{3}{c|}{Gaussian Noise} 
        & \multicolumn{3}{c}{Skewed Noise} \\
        & \textbf{NN} & \textbf{LS} & \textbf{LS Idealized}
        & \textbf{NN} & \textbf{LS} & \textbf{LS Idealized} \\
        \hline
    
        $d_0$ & $0.0076\pm0.01$ & $0.013\pm0.1$ & $0.0093\pm0.01$ & $0.014\pm0.02$ & $0.13\pm0.2$ & $0.28\pm0.3$ \\
        $\phi_0$ & $0.0035\pm0.005$ & $0.005\pm0.04$ & $0.0038\pm0.005$ & $0.015\pm0.02$ & $0.069\pm0.08$ & $0.038\pm0.05$ \\
        $1/p_\mathrm{T}$ & $0.00032\pm0.0004$ & $0.00047\pm0.004$ & $0.00033\pm0.0004$ & $0.0018\pm0.002$ & $0.0055\pm0.006$ & $0.0032\pm0.004$ \\
        $dz$ & $0.0056\pm0.007$ & $0.0062\pm0.04$ & $0.0055\pm0.007$ & $0.049\pm0.06$ & $0.3\pm0.07$ & $0.3\pm0.07$ \\
        $\tan(\lambda)$ & $0.00096\pm0.001$ & $0.001\pm0.03$ & $0.00088\pm0.001$ & $0.0079\pm0.01$ & $0.0089\pm0.01$ & $0.0089\pm0.01$ \\
        \hline\hline
    \end{tabular}
\end{table}

In the Gaussian noise model, both methods show no bias, as expected, and the NN matches the performance of the idealized LS method, which suggests that it is approaching the optimal precision.  Note that the $\chi^2$ of tracks from the LS methods are consistently lower than that of the true parameters, as expected from a method which explicitly minimizes $\chi^2$ in data which includes noise that shifts the minimum from the true location. The NN method, by contrast, produces tracks whose $\chi^2$ distribution is similar to that of the true parameters, as expected from a method which minimizes the parameter residual rather than the $\chi^2$.  In the case of Gaussian noise, these two optimization targets yield similar performance in parameter residuals, as expected.


In the skewed noise model, the difference in the approaches becomes apparent. Here, the LS method fails to provide an unbiased estimate, because its assumed Gaussian noise model does not accurately describe the data. In scenarios where hits have fluctuated in the same direction from the true track, the LS model prefers to follow the hits at the expense of the parameter residuals. The NN model  learns the implicit relationship between the hit positions and the track parameters, without any requirement that it be explicitly specified.

In the presence of the larger skewed noise, the LS model becomes more likely to get stuck in local minima. This can be seen in the track parameter residuals in Fig \ref{fig:residuals}. The distributions of $d_0,\phi_0, 1/p_{\textrm{T}}$ appear bimodal, indicating alternative minima. When comparing the idealized and realistic LS setups, the primary distinction lies in the starting parameters: the idealized case begins at the true parameters plus noise, while the realistic case starts from the fast-fit initial guess. The idealized approach almost always begins in the correct basin of attraction and thus avoids convergence to a local minimum. In contrast, an imperfect fast fit can place the initial guess in the wrong loss-function valley, leading to convergence on an alternative local minimum. When substantial noise is present, such minima can yield nearly identical $\chi^2$ values to the true helical-parameter minimum, making it difficult for the fast fit to consistently select the correct basin. In the Skewed case, the $\chi^2$ plots comparing realistic and idealized LS are still nearly identical to each other despite the differences in predicted helical parameters, see Fig. \ref{fig:chisq} and Fig. \ref{fig:chisq_truth}. This also explains why, for the Gaussian (lower-noise) data, the idealized and realistic methods are both unimodal.

If one knew the true noise model, it is in principle sufficient to modify the likelihood to describe it, though a well-behaved likelihood would require a smooth noise to ensure convergence. The NN requires no explicit assumed noise model, neither closed form nor continuous, only that used implicitly to generate the training sample.  In that sense, it of course still requires a noise model, but only in the training sample, without the limitations of expressing a well-behaved version in the likelihood. This allows for a broader set of noise models, including those which are bi-modal or not smooth, or measured directly in test-beam data.

Here, the deviation of the skewed noise from the Gaussian represents an exaggerated example of a typical scenario, where the need for a smooth and simple noise model in the least-squares fit fails to capture the true distribution of  hit locations. The LS model does achieve smaller values of $\chi^2_{\textrm{hit}}$, which demonstrates that in scenarios where the assumed noise model does not match the actual noise, $\chi^2_{\textrm{hit}}$ is no longer a proxy for the true target, the helix parameters.

\clearpage
\section{Anomaly Detection}
\label{sec:ad}

Particle tracking is a vital aspect of the reconstruction of well-known particles, such as electrons or muons, but also allows for discovery of new particles whose trajectories do not follow helical paths~\cite{Kang:2008ea, Sha:2024hzq}. The broadest search for such particles would rely on anomaly detection, searches for tracks that deviate from the helical, rather than hypothesis testing, searches for one specific class of new track.  

A well-known strategy for anomaly detection is via use of a latent space. An encoder is trained to map data into the latent space, and a second decoder learns to map back to the data space and the pair are encouraged to minimize the difference between the original and mapped data, the reconstruction error.  When out-of-distribution data which was not present in the training set is mapped, it receives a large reconstruction error. 

Identifying the helical parameters as the latent space, a sort of compression of the hits into a helical hypothesis, allows for a natural application of our learned $f(\bar{x})$ and analytical $h(\bar{p})$ in the context of anomaly detection.   Tracks that are well described by a helix can be accurately reconstructed (via the decoder $h(\bar{p})$) from their parameters. Tracks whose hits are not helical will be compressed into a set of helical parameters in the latent space, but the track extrapolated from those parameters will not match the original hits, giving a large reconstruction loss.  Hence, learning $f(\bar{x})$ allows us to pair it with $h(\bar{p})$ to serve as an anomaly detection algorithm.

Of course, existing LS method already provides a goodness-of-fit test via  $\chi^2_\textrm{hit}$. But if $f(\bar{x})$ is a more accurate encoder, it can serve as the first element of a more powerful anomaly detector. In addition, its flexibility in the noise model allows us to avoid false positives more effectively.

\begin{figure}
    \includegraphics[width=0.45\linewidth]{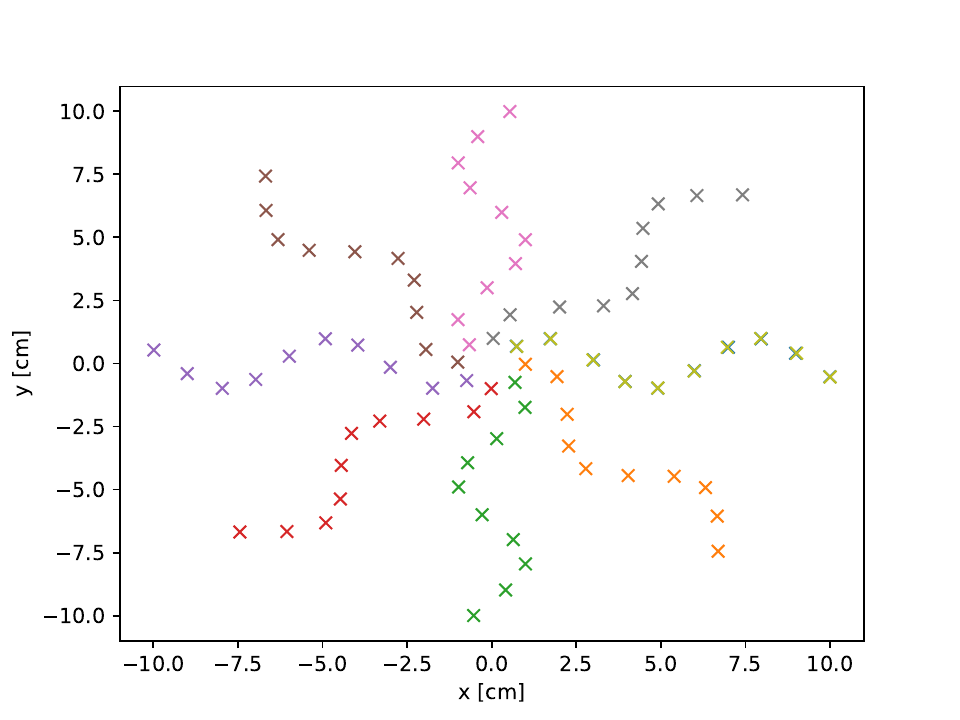}
    \includegraphics[width=0.45\linewidth]{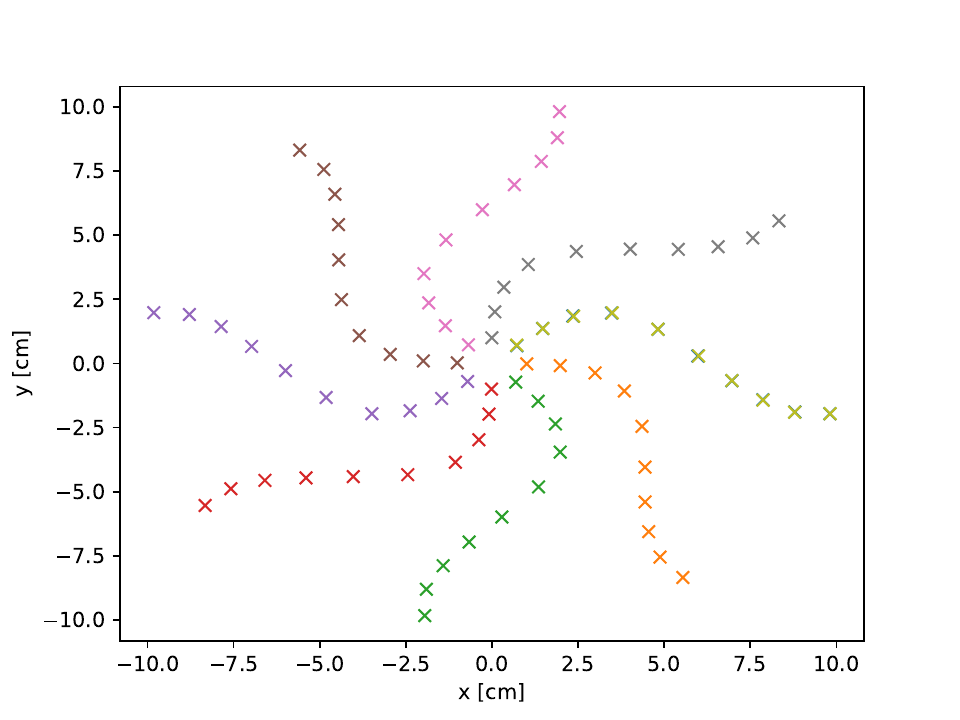}
        \caption{Examples of non-helical generated tracks with varying initial angle for two values of $A,\omega$.}
        \label{fig:gentracks_sin}
\end{figure}

We generate a sample of 100K non-helical tracks, and combine with 100K helical tracks to create a classification dataset. Inspired by the quirk~\cite{Kang:2008ea} trajectories, we create tracks which follow a sine curve in the $xy$ plane, described by a frequency $\omega$ and an amplitude $A$, rotated into the angle $\phi_0$. Parameter ranges are given in Tab. ~\ref{tab:non_helical_track_params} and example tracks are shown in Fig.~\ref{fig:gentracks_sin}.

\begin{table}[htbp]
  \centering
  \begin{tabular}{lcc}
    \hline
    \textbf{Parameter} & \textbf{Distribution} & \textbf{Range} \\
    \hline
    $d_0$   & Half-Normal ($\sigma=0.01$) & $[0,\infty)$ \\
    $\phi$  & Uniform  & $[0,2\pi]$ \\
    $A$     & Uniform  & $[0.5,1.5]$ \\
    $\omega$ & Uniform & $[0.5,3.0]$ \\
    $dz$    & Normal ($\sigma=1.0$) & $(-\infty,\infty)$ \\
    $\tan\lambda$ & Normal ($\sigma=0.3$) & $(-\infty,\infty)$ \\
    \hline
  \end{tabular}
  \caption{Ranges and distributions of the non-helical track parameters.}
  \label{tab:non_helical_track_params}
\end{table}

\begin{figure}
    \centering
    \includegraphics[width=0.9\linewidth]{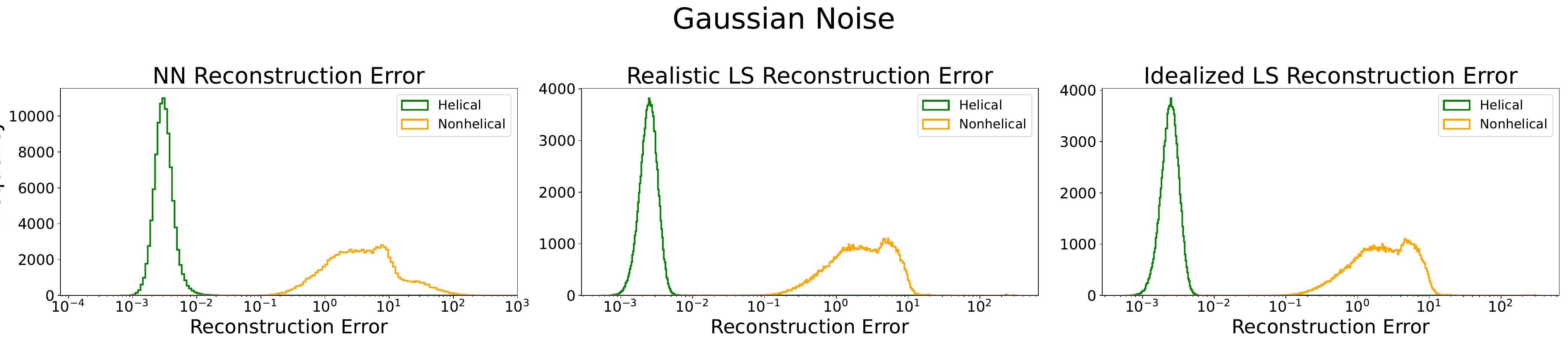} 
    \includegraphics[width=0.9\linewidth]{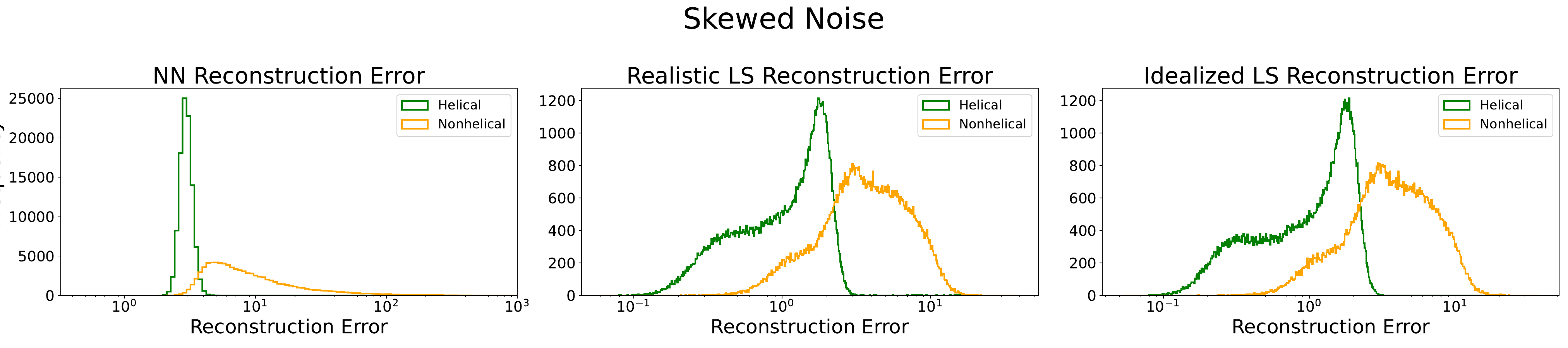}

    \caption{ Demonstration of track parameter estimation as the first step in an anomaly detection pipeline, which maps the hits to the parameters and back to the hit space. The reconstruction score is $\chi^2_{hit}$, the distance between the observed and reconstructed hits. Shown are distributions of the score for helical and non-helical tracks for the NN-based (left) and LS-based (right) pipelines, under the Gaussian (top) and Skewed (bottom) noise models.}
    \label{fig:AD}
\end{figure}

We use the learned $f(\bar{x})$, trained only on helical tracks, together with the helical parametrization function $h(\bar{p})$ to distinguish between helical and non-helical tracks.  Figure~\ref{fig:AD} shows distributions of the resulting reconstruction loss in the Gaussian and Skewed noise models. 

In the Gaussian case, both the NN and LS approaches are very effective (each achieves over 99\% accuracy), as the non-helical tracks are easily distinguishable from the helical tracks. The current standard deviation used in the gaussian noise is 0.01, but at higher levels of noise, the problem can be made increasingly more difficult. For example, with $\sigma=0.2$, there are many helical tracks which mimic the behavior of a non-helical trajectory. At higher levels of noise, the task of anomaly detection becomes more difficult.

In the skewed case, the task is more difficult, and here the NN does significantly better than the LS pipeline. The NN has classification accuracy of 96.2\%  versus 85.6\% for the LS pipeline.

\begin{figure}
    \centering
    \includegraphics[width=0.4\linewidth]{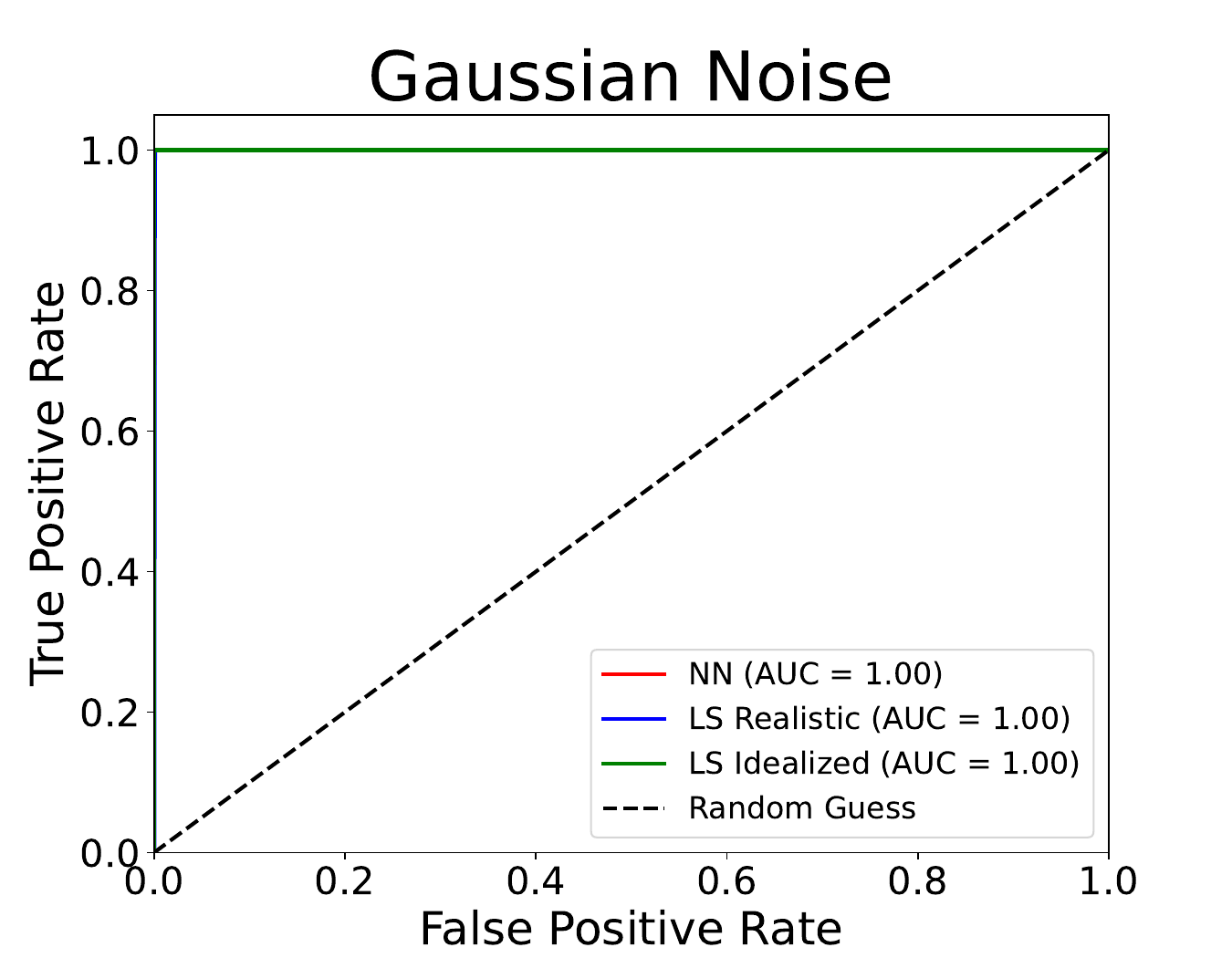} 
    \includegraphics[width=0.4\linewidth]{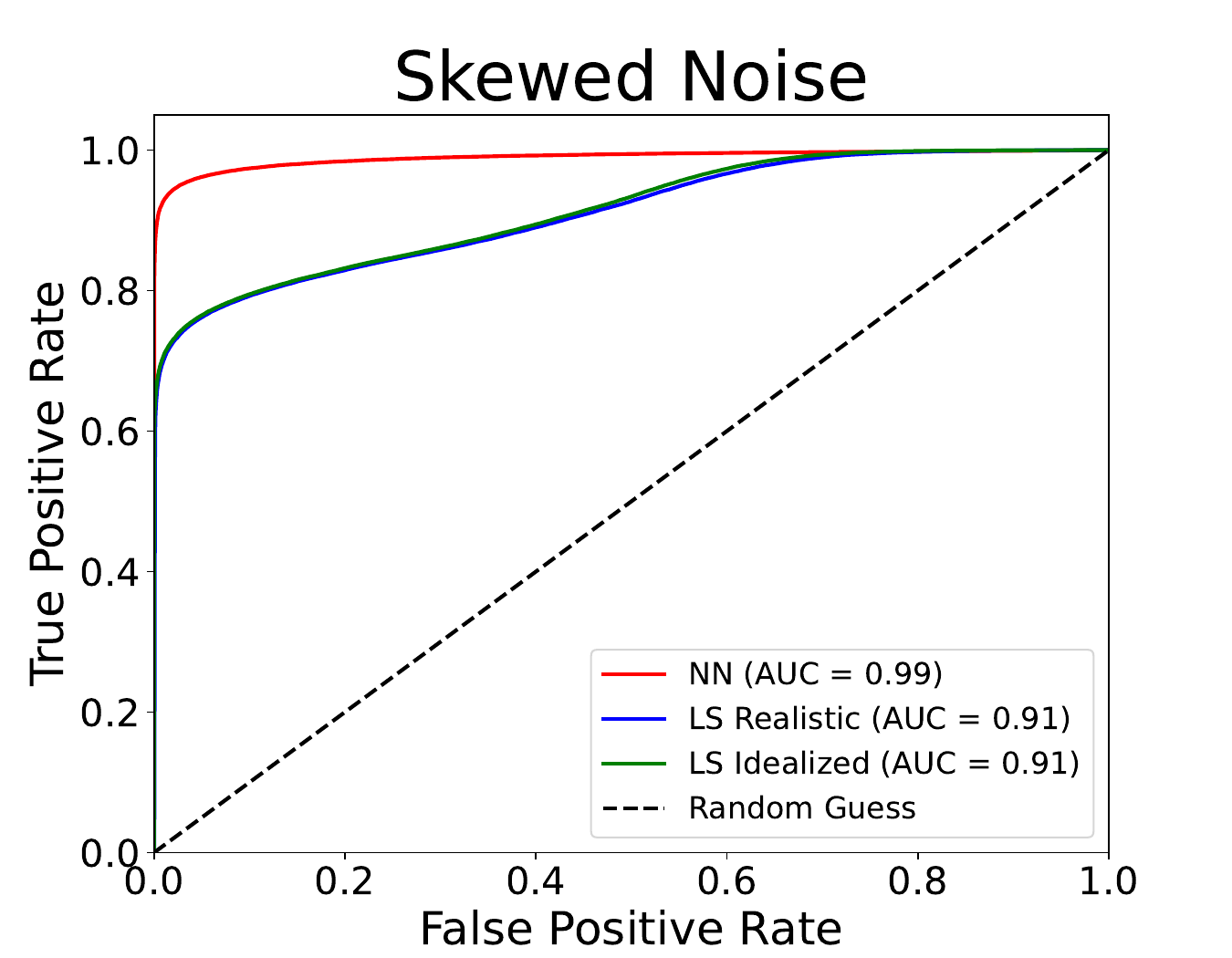}
    \caption{ True versus false positive rate for track anomaly detection of non-helical tracks using autoencoders built with the NN track fitter or the LS method. Left is for the Gaussian noise model, right for the skewed case.}
    \label{fig:ROC}
\end{figure}

\section{Conclusions}
\label{sec:conc}

The task of track fitting is to most accurately, and without bias, recover the parameters of the helical trajectory of a charged particle. These parameters allow for the extraction of crucial quantities such as muon momentum, or interaction vertex locations. Any improvement, such as a reduction in the variance of the parameters, could have broad impact on the power of particle physics experiments to make precise measurements of Standard Model parameters or discover new physics. 
However, in its traditional form inspired by maximal likelihood techniques, track fitting presents a significant computational challenge due to the need to search the helical parameter space.

Here, we have presented an alternative approach using  neural network regression to map the 3D space points of the hits directly to the most likely helical parameters. By casting fitting as a machine learning task, we shift the computational cost from per-track optimization to learning a hit-to-parameter mapping. This also allows for optimization of the true goal: extraction of the helical parameters.  When compared to the traditional, least-squares approach which searches for the track that minimizes the distance to the hits, our trained track fitter provides low variance in the track parameter residuals, can naturally accommodate non-Gaussian uncertainties without introduction of bias, and operates over 1,000 times faster without use of stochastic optimization heuristics.  Potential applications of accurate, robust to non-Gaussian-noise, computationally cheaper track fitting include improved track finding in algorithms such as Kalman filters~\cite{FRUHWIRTH1987444} that iterate between finding and fitting, rapid tracking in the trigger setting, improved estimation of muon momenta, more accurate discrimination between electron and photon candidates, identification and recovery of Bremsstrahlung effects on electron candidates, and more accurate displaced vertex and jet substructure measurements.

The work presented here provides a point estimate of the parameters. The downstream applications discussed above, however, additionally rely on information from the fit covariance matrices that encode information on uncertainties and correlations between the estimated parameters. A covariance matrix can still be obtained in our approach through traditional least-square methods around the neural network fit. Although, as this approach associates covariances obtained through minimizing $\chi^2_\textrm{hit}$ with parameters obtained through minimizing track parameter residuals, it is naturally more representative in the Gaussian uncertainty limit. Future work can explore the use of generative models including variational autoencoders~\cite{kingma2022autoencodingvariationalbayes} or normalizing flows to learn the full posterior distribution over the parameters. This generative approach could provide a more robust estimate of uncertainty due and account for situations where observation noise leads to more than one viable parametrization. In addition, this study is a proof-of-principle demonstration in a simplified environment that includes a noise model, but does not have the realism of a GEANT4-based simulation~\cite{AGOSTINELLI2003250}. Missing and merged hits, for example, have not been considered, though in principle the model is flexible enough to learn these if they are present in the training samples. More broadly, by learning from examples, our approach can accommodate any features, such as missing and merged hits and different kinds of noise, that can be realistically incorporated in the training data.

\section*{Acknowledgements}
The authors are grateful to Aishik Ghosh for helpful comments. MV, LC and DW are supported by the DOE Office of Science.

\clearpage
\appendix

\section{  Geometry of the data and parameter spaces.}

We search for the parameters $\bar{p}$ which minimize  $\bar{p}-\bar{p}_0$, the distance to the true parameters $\bar{p}_0$.  Our committee-of-experts approach allows us to train each network to optimize for a single parameter, requiring no assumptions about the structure of the 5-dim space and how one might balance residuals in each dimension.

\begin{figure}[h!]
    \centering
  \includegraphics[width=0.4\linewidth]{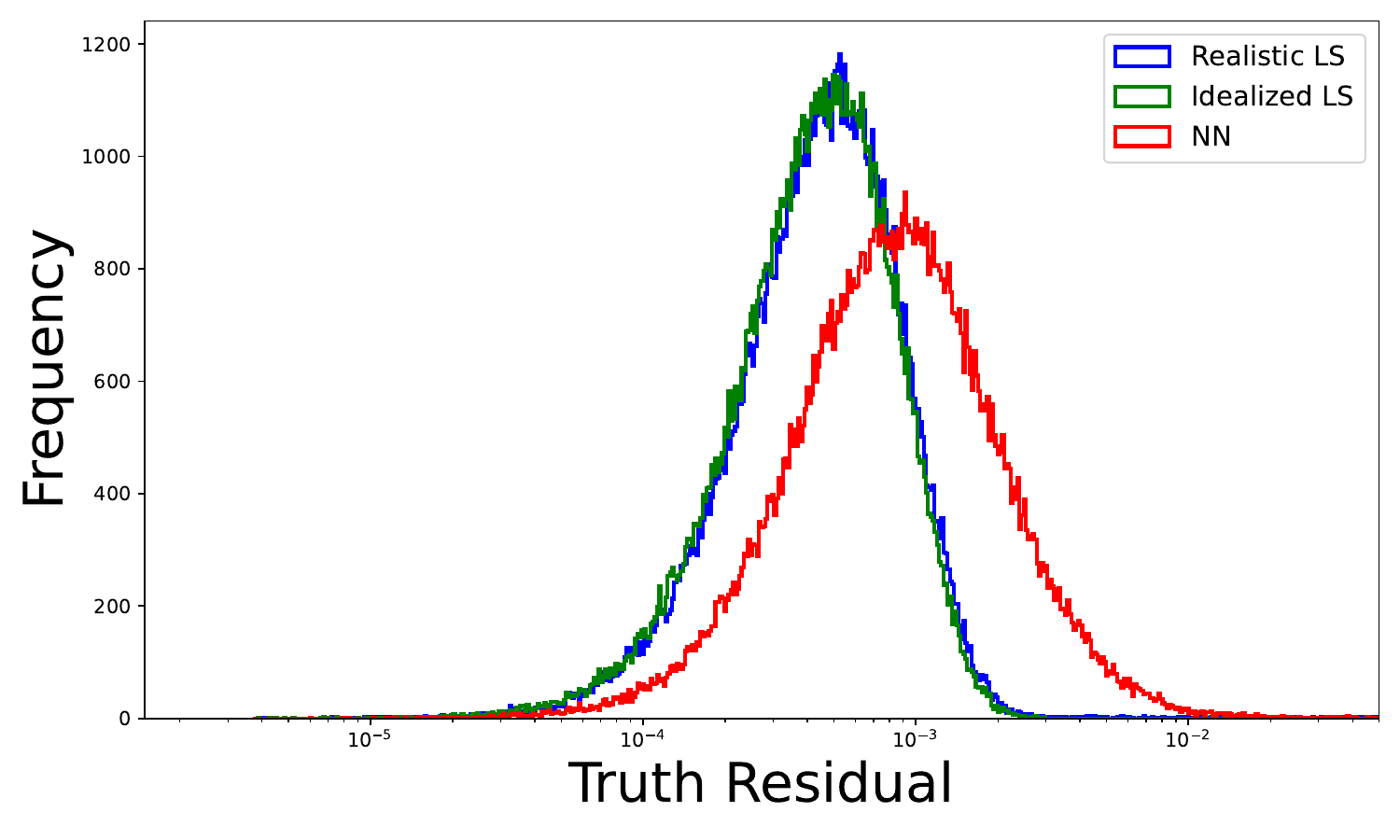} 
    \includegraphics[width=0.4\linewidth]{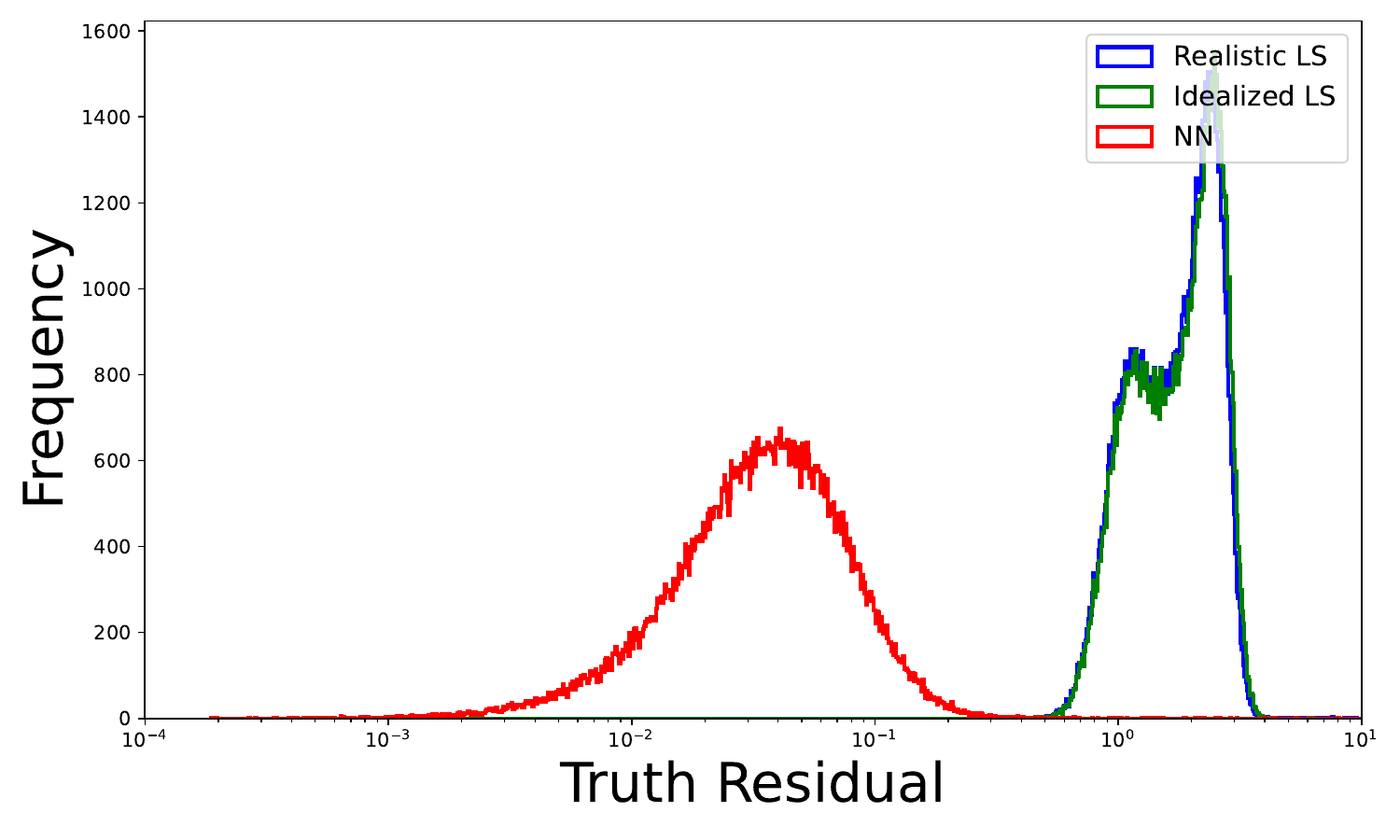}
    \caption{Distribution of the $\chi^2_\textrm{truth}$ quantity for both models. Left, Gaussian noise, right skewed. }
    \label{fig:chisq_truth}
\end{figure}

One could also consider mapping to the data space, training a network to minimize 
\[\chi^2_{\textrm{truth}} = \frac{[h(\bar{p})-h(\bar{p}_0)]^2 }{\sigma^2} = \frac{[\bar{x}-h(\bar{p}_0)]^2 }{\sigma^2}  \] 

\noindent which is in principle equivalent to minimizing $\bar{p}-\bar{p}_0$, but conceptually more similar to the traditional approach which minimizes the hit-to-track distance. Distributions of the $\chi^2_\textrm{truth}$ quantity for both approaches
are shown in Fig.~\ref{fig:chisq_truth}. This hit-space likelihood would prioritize reconstructing the tracks, down-weighting the importance of parameters which would not affect the overall hit locations. For example, the effect of the transverse momentum parameter is less important for tracks with a smaller deviation since the overall helix radius will be smaller. Additionally, due to observation noise, one set of observed hits may actually be explained by more than one distinct set of track parameters. Enforcing a single truth-value parameter is therefore not theoretically sound and may destabilize training in a more complex scenario with unknown noise models. This approach, however, introduces numerically unstable gradients due to the non-linear nature of the helix in the $x-y$ plane, and requires assuming a noise model.

\section{Implicit weights on track parameters in least squares fitting}
\label{app:chisq}

To demonstrate the complex relationships between the parameter and data spaces, we study the impact on $\chi^2_{fit}$ when track parameters are shifted. See Tab~\ref{tab:chisq_experiment} 

\begin{table}[h]
    \centering 
    \caption{Differences in hit residual value when true helical parameter $p_i$ has a small perturbation of size 0.01 added to it. The baseline is the hit residuals when all the helical parameters are equal to the true target parameters. This is nonzero due to the Gaussian noise.}
    \label{tab:chisq_experiment}
    \begin{tabular}{c|c}
        \hline\hline
        \textbf{Helical Parameter} & \textbf{Mean Hit Residual Difference} \\
        \hline
        \(d_0\) & 0.00098 \\
        \hline
        \(\phi_0\) & 0.039 \\
        \hline
        \(1/p_T\) & 2.73 \\
        \hline
        \(dz\) & 0.0010 \\
        \hline
        \(\tan(\lambda)\) & 0.039 \\
        \hline
         Baseline & 0.0030 \\
        \hline\hline
    \end{tabular}
\end{table}

\section{ Prediction speed vs number of hits}

In order to provide a more detailed assessment of model speed, we evaluate the prediction speed of both the LS and NN methods as the number of hits per track is increased from the default value of 10. We evaluate the NN using a batch size of 100k. See Tab~\ref{tab:speedup_experiment} 

\begin{table}[h]
    \centering 
    \caption{Prediction speed of NN and LS models as number of hits per track is varied.}
    \label{tab:speedup_experiment}
    \begin{tabular}{c|c|c}
        \hline\hline
        \textbf{Number of Hits} & \textbf{NN Speed (s/track)} & \textbf{LS Speed (s/track)}\\
        \hline
        10 hits & \(0.000208\) & 0.268\\
        \hline
        20 hits & \(0.000212\) & 0.269\\
        \hline
        30 hits & \(0.000212\) & 0.272 \\
        \hline
        40 hits & \(0.000213\) & 0.271 \\
        \hline\hline
    \end{tabular}
\end{table}

Although there is a slight increase between trials, the inference time of both the NN and LS methods appear relatively constant as the number of hits per track is scaled up.

\section{ Model bias from parameter distribution}

One downside with the neural network approach is that the MSE loss can cause a bias in the model predictions. Because each training point is weighted equally and we are trying to minimize the total parameter loss, the model may focus on predicting regions of parameter space that we sample most densely. We tweak the parameter distributions to create a test set of 100K points with differently distributed parameters. The new distributions can be seen in Tab~\ref{tab:track_params}

\begin{table}[htbp]
  \centering
  \begin{tabular}{lcc}
    \hline
    \textbf{Parameter} & \textbf{Distribution} & \textbf{Range} \\
    \hline
    $d_0$   & Half-Normal ($\sigma=0.01$)  & $[0,\infty)$ cm \\
    $\phi$  & Uniform   & $[0,2\pi]$ rad  \\
    $p_\textrm{T}$   & Uniform   & $[25,200]$ GeV  \\
    $dz$    & Normal ($\sigma=1.0$) & $(-\infty,\infty)$ cm\\
    $\tan(\lambda)$ & Normal ($\sigma=0.3$) & $(-\infty,\infty)$ \\
    \hline
  \end{tabular}
  \caption{Ranges and distributions of the helical track parameters.}
  \label{tab:track_params}
\end{table}

We predict helical parameters on this modified test set, and compare the mean absolute errors between the modified test set and the Gaussian test set in Tab~\ref{tab:helical_parameters_alt_distrubtion}

\begin{table}[h!]
    \centering
    \caption{Mean absolute errors with standard deviations of helical parameters for Gaussian and modified test sets.}
    \label{tab:helical_parameters_alt_distrubtion}
    \begin{tabular}{c|cc|cc}
        \hline\hline
        \multirow{2}{*}{\textbf{Parameter}} 
        & \multicolumn{2}{c|}{Gaussian Test Set} 
        & \multicolumn{2}{c}{Modified Test Set} \\
        & \textbf{NN} & \textbf{LS}
        & \textbf{NN} & \textbf{LS} \\
        \hline
    
        $d_0$ & $0.0076\pm0.01$ & $0.013\pm0.1$ & $0.0072\pm0.007$ & $0.01\pm0.05$ \\
        $\phi_0$ & $0.0035\pm0.005$ & $0.005\pm0.04$ & $0.0037\pm0.004$ & $0.004\pm0.009$ \\
        $1/p_\mathrm{T}$ & $0.00032\pm0.0004$ & $0.00047\pm0.004$ & $0.00038\pm0.0005$ & $0.00035\pm0.0005$ \\
        $dz$ & $0.0056\pm0.007$ & $0.0062\pm0.04$ & $0.0061\pm0.008$ & $0.0055\pm0.01$ \\
        $\tan(\lambda)$ & $0.00096\pm0.001$ & $0.001\pm0.03$ & $0.0013\pm0.003$ & $0.00093\pm0.01$ \\
        \hline\hline
    \end{tabular}
\end{table}

The NN model experiences an increase in parameter residuals on the modified test set, indicating some small biases learned from the training data. Alternatively, the LS model experiences decreases in parameter residuals, which may indicate the modified test set is slightly easier to fit. Regardless, both models are still able to accurately fit the tracks, and the differences between the test sets are marginal.

\section{ Helical parameter pair plots }

We provide pair plots between parameter residuals for the 3 models to understand the correlations between helical parameters.

\begin{figure}
    \centering
    \includegraphics[width=0.9\linewidth]{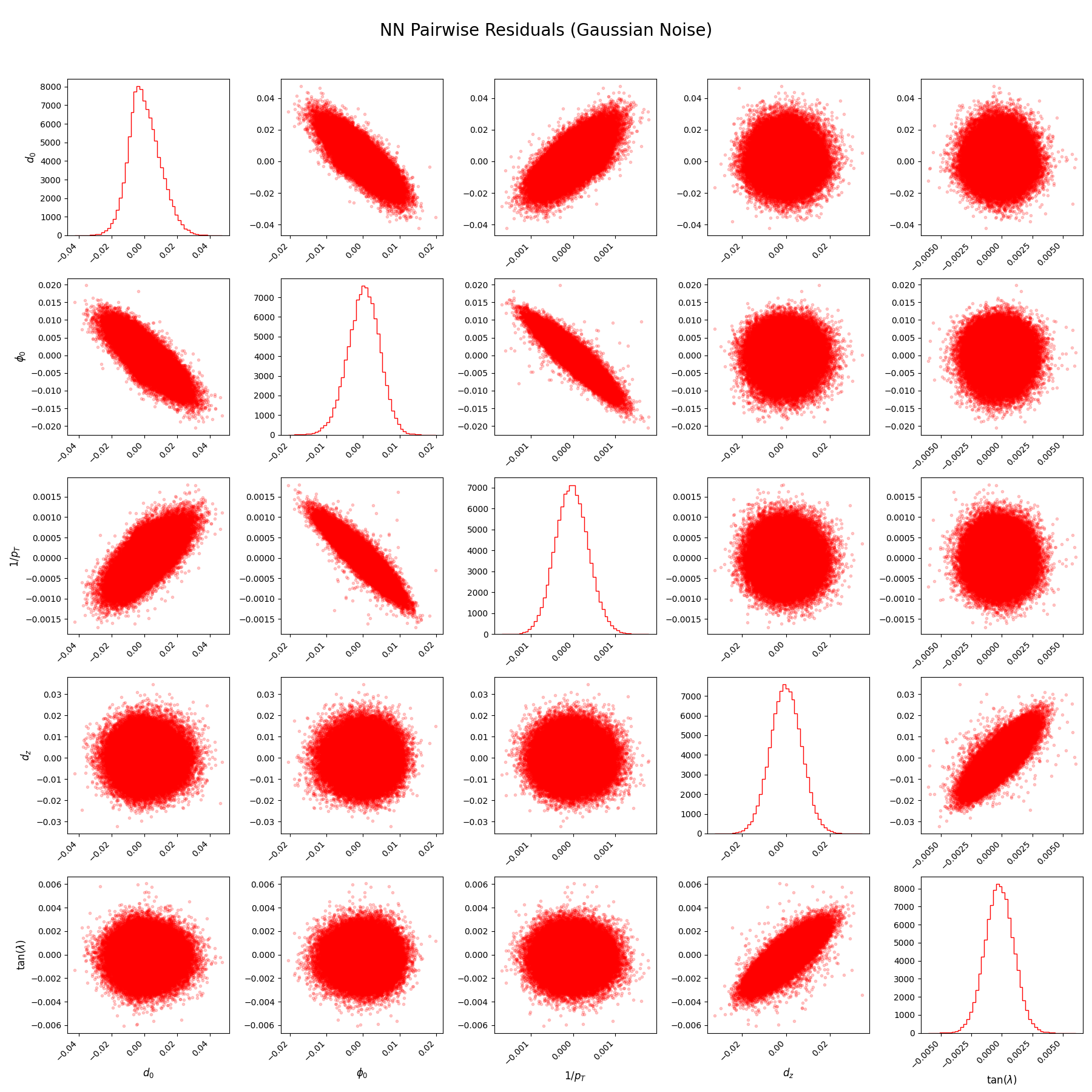}
    \caption{ NN parameter residual pair plot on Gaussian dataset.}
    \label{fig:nn_gaussian_pair}
\end{figure}

\begin{figure}
    \centering
    \includegraphics[width=0.9\linewidth]{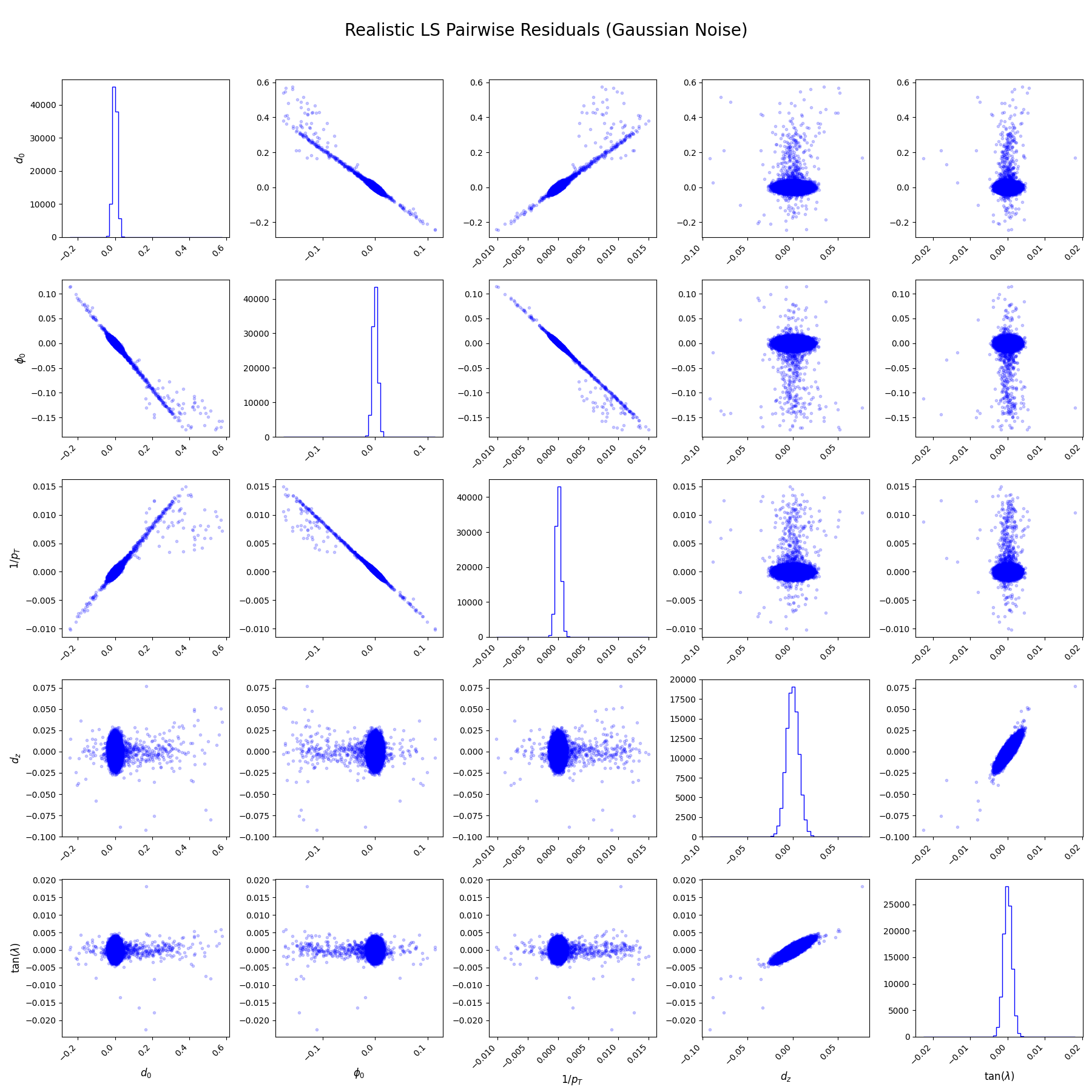}
    \caption{ Realistic LS parameter residual pair plot on Gaussian dataset.}
    \label{fig:realistic_ls_gaussian_pair}
\end{figure}

\begin{figure}
    \centering
    \includegraphics[width=0.9\linewidth]{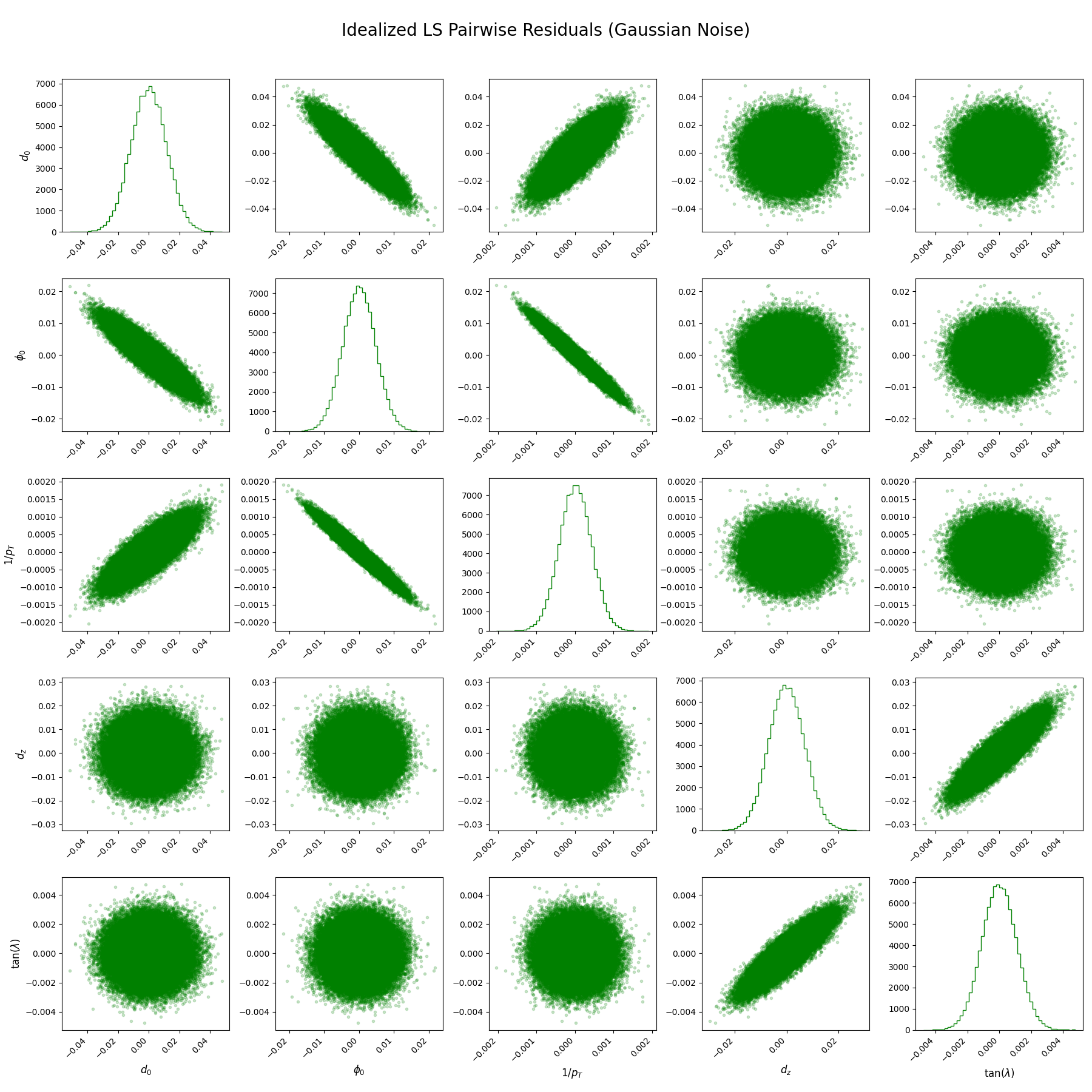}
    \caption{ Idealized LS parameter residual pair plot on Gaussian dataset.}
    \label{fig:idealized_ls_gaussian_pair}
\end{figure}

\begin{figure}
    \centering
    \includegraphics[width=0.9\linewidth]{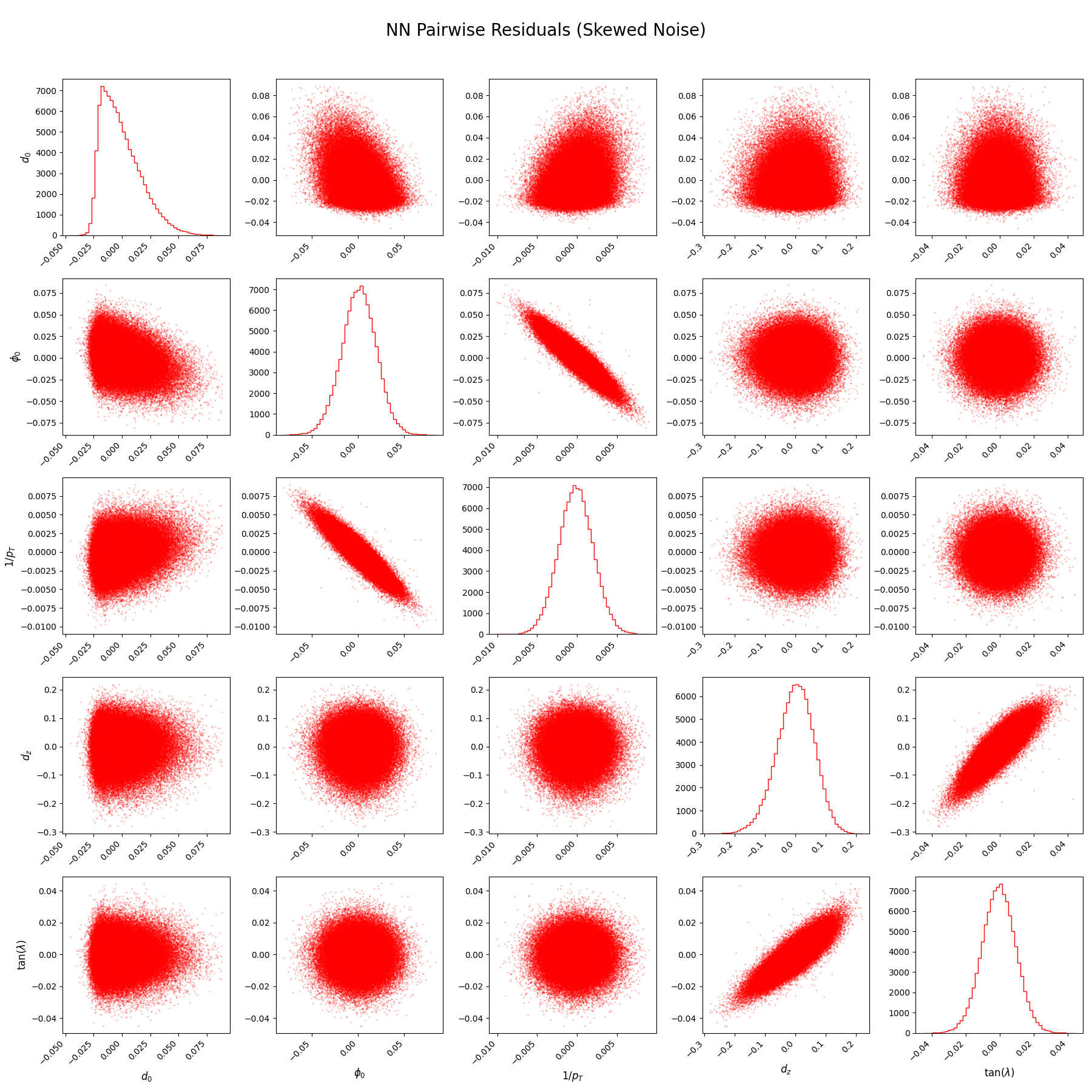}
    \caption{ NN parameter residual pair plot on Skewed dataset.}
    \label{fig:nn_skewed_pair}
\end{figure}

\begin{figure}
    \centering
    \includegraphics[width=0.9\linewidth]{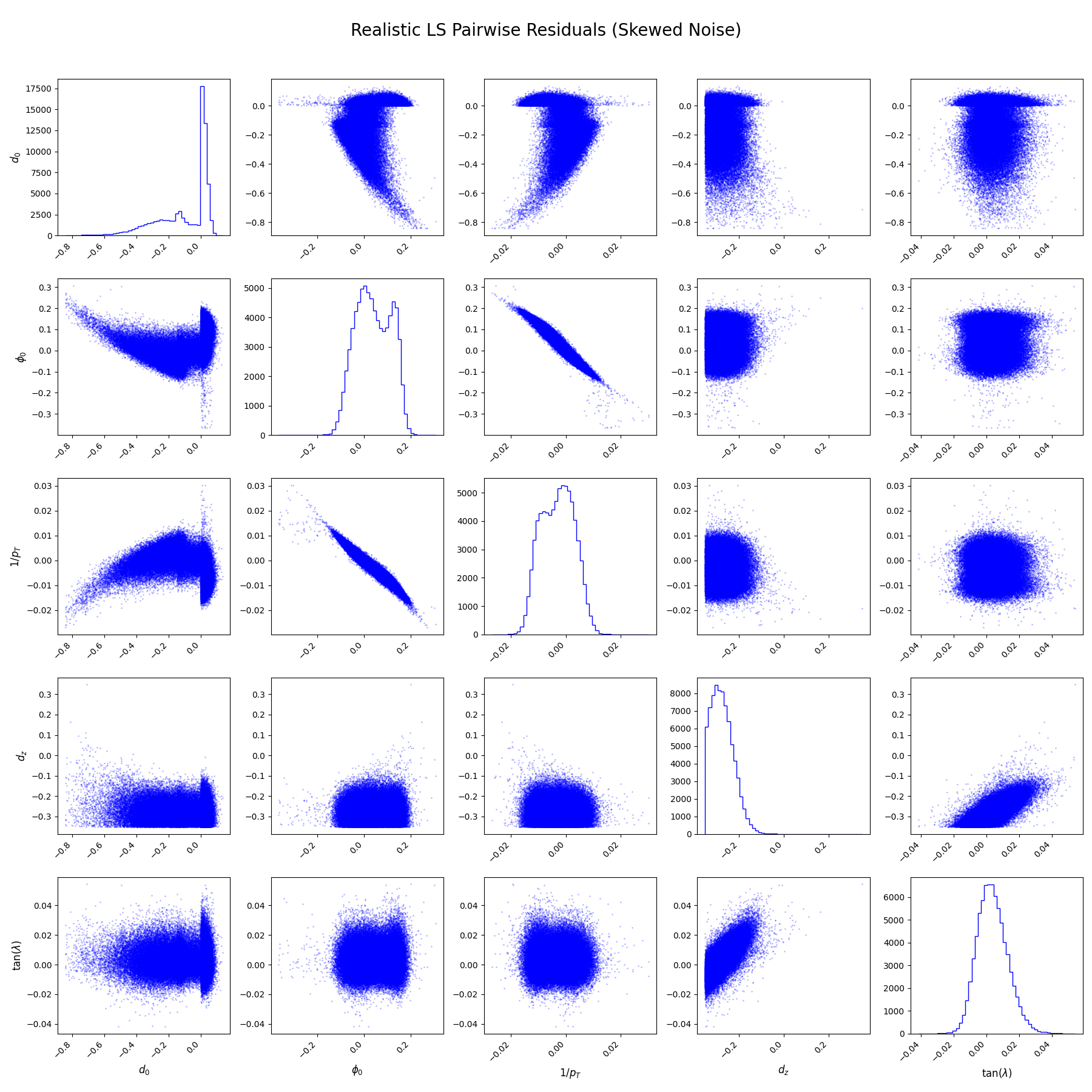}
    \caption{ Realistic LS parameter residual pair plot on Skewed dataset.}
    \label{fig:realistic_ls_skewed_pair}
\end{figure}

\begin{figure}
    \centering
    \includegraphics[width=0.9\linewidth]{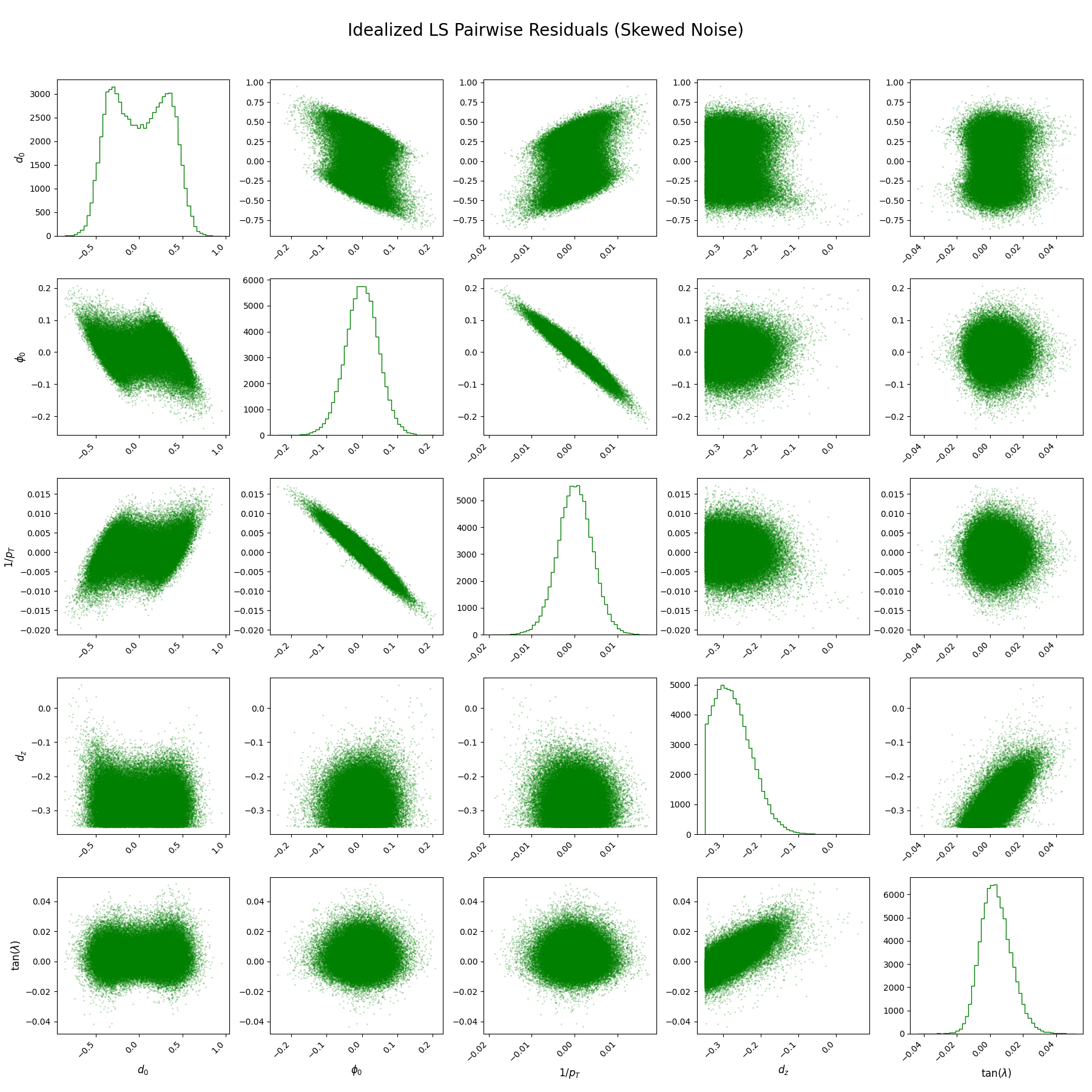}
    \caption{ Idealized LS parameter residual pair plot on Skewed dataset.}
    \label{fig:idealized_ls_skewed_pair}
\end{figure}

\bibliography{tracks}

\end{document}